\documentclass[iop]{emulateapj}

\newcommand{\arcm}{\ifmmode {' }\else $' $\fi}
\newcommand{\arcs}{\ifmmode {'' }\else $'' $\fi}

\newcommand{\lapp}{$_<\atop{^\sim}$}
\newcommand{\gapp}{$_>\atop{^\sim}$} 
\newcommand{\ngc}{$N_{\rm GC}$}
\newcommand{\mbh}{$M_{\rm SMBH}$}
\slugcomment{}

\shortauthors{Rhode} \shorttitle{Globular Cluster-Supermassive Black Hole Relation}

\begin{document}

\title{Exploring the Correlations between Globular Cluster
  Populations and Supermassive Black Holes in Giant Galaxies} 

\author{Katherine L. Rhode\altaffilmark{1}}
\affil{Department of Astronomy, Indiana University, 727 East Third
  Street, Bloomington, IN 47405; rhode@astro.indiana.edu} 

\altaffiltext{1}{Visiting Astronomer, Kitt Peak National Observatory,
National Optical Astronomy Observatory, which is operated by the
Association of Universities for Research in Astronomy (AURA)
under cooperative agreement with the National Science Foundation.}

\begin{abstract}
This paper presents an analysis of the correlation between the number
of globular clusters (\ngc) in giant galaxies and the mass of the
galaxies' central supermassive black hole (\mbh).  I construct a
sample of 20 elliptical, spiral, and S0 galaxies with known SMBH
masses and with accurately-measured globular cluster system properties
derived from wide-field imaging studies.  The coefficients of the
best-fitting \ngc$-$\mbh\ relation for the early-type galaxies are
consistent with those from previous work but in some cases have
smaller relative errors.  I examine the correlation between \ngc\ and
\mbh\ for various subsamples and find that elliptical galaxies show
the strongest correlation while S0 and pseudobulge galaxies exhibit
increased scatter.  I also compare the quality of the fit of the
numbers of metal-poor globular clusters versus SMBH mass and the
corresponding fit for metal-rich globular clusters.  I supplement the
20-galaxy sample with ten additional galaxies with reliable
\ngc\ determinations but without measured \mbh. I use this larger
sample to investigate correlations between \ngc\ and host galaxy
properties like total galaxy luminosity and stellar mass and bulge
luminosity and mass.  I find that the tightest correlation is between
\ngc\ and total galaxy stellar mass.  This lends support to the notion
that \ngc\ and \mbh\ are not directly linked but are correlated
because both quantities depend on the host galaxy potential.  Finally,
I use the \ngc$-$\mbh\ relation derived from the 20-galaxy sample to
calculate predicted \mbh\ values for the ten galaxies with accurate
\ngc\ measurements but without measured SMBH masses.
\end{abstract}


\section{Introduction}
\label{section:introduction}

It is now well-known that a correlation exists between the mass of the
supermassive black hole (SMBH) at the centers of giant
galaxies and the velocity dispersion of the galaxies' spheroidal
component
(Ferrarese \& Merritt 2000, Gebhardt et al.\ 2000b and many subsequent
papers).  This ``$M$-$\sigma$ relation'' has been interpreted as
indicating a close connection between the formation and evolution of
galaxy bulges and central SMBHs.  Much observational and theoretical
work has been devoted to establishing the exact form of the relation
(e.g., Tremaine et al.\ 2002, Gultekin et al.\ 2009) as well as the
reasons for its existence (e.g., Burkert \& Silk 2001, Miralda-Escude
\& Kollmeier 2005, Di Matteo et al.\ 2005, Robertson et al.\ 2006,
Johansson et al.\ 2009, Jahnke \& Maccio 2011). The general consensus
seems to be that the
presence of SMBHs at the centers of giant galaxies is a natural
consequence of hierarchical structure formation and fits well within
our overall picture of galaxy assembly (e.g., Robertson et al.\ 2006,
Jahnke \& Maccio 2011).  We also know that SMBHs must have formed
relatively early in many galaxies' histories: observations of
high-redshift quasars demonstrate that SMBHs existed at $z$ of
$\sim$6$-$7 (e.g., Fan et al.\ 2006, Mortlock et al.\ 2011),
when the Universe was less than a Gigayear old.  On the other hand,
the detailed physical mechanisms responsible for seeding the central
SMBHs in the first place, and then growing them over time, are not yet
well-understood (e.g., Volonteri \& Rees 2006, Omukai, Schneider, \&
Haiman 2008, Mayer et al.\ 2010).

Like SMBHs, many globular clusters (GCs) apparently also formed during
the early stages of galaxy formation.  GCs are luminous ($M_V$ $-$11
to $-$4), compact ($R_{1/2}$ of a few parsecs), populous
($\sim$$10^4-10^6$ members) star clusters that orbit their host
galaxies at galactocentric distances of less than a kiloparsec out to
hundreds of kiloparsecs (e.g., Rhode \& Zepf 2004, Rhode et
al.\ 2007).  GCs in the Milky Way, which arguably have the most
accurate absolute age determinations, typically have ages of
$\sim$11$-$13~Gyr (e.g., Dotter et al.\ 2010, Forbes \& Bridges 2010,
and references therein).  Observations of external galaxies have
detected GCs with intermediate ages (e.g., 1.5$-$4 Gyr; Goudfrooij et
al.\ 2007) as well as young ($<$1~Gyr) star clusters, with masses
equal to or greater than GCs (e.g., Bastian et al.\ 2006, Whitmore et
al.\ 2010), that were likely produced in gas-rich galaxy mergers.
Because they appear to have formed during intense star formation
events, including events triggered by
mergers, the properties of GCs can be used to trace the major assembly
and evolutionary episodes of their host galaxies \citep{az98,
  brodie06}.  The total numbers, spatial distributions, ages,
metallicities, and kinematics of GCs provide important physical clues
regarding the origin and evolution of their parent galaxies.

A few recent papers have begun to explore possible links between these
two constituents of galaxies, SMBHs and GCs.  Spitler \& Forbes (2009)
presented a method for using the total number of GCs (\ngc) in a
galaxy to estimate the mass of the galaxy's halo.  They in turn showed
that, for a sample of about a dozen galaxies, the halo mass derived
from the number of GCs correlates fairly well with the measured SMBH
mass.  Burkert \& Tremaine (2010; hereafter BT10) investigated a more
direct correlation between SMBHs and GCs by showing that there is a
tight correlation between \ngc\
and the mass of the central SMBH (\mbh) in early-type galaxies.  Using
a sample of 13 galaxies with \ngc\ estimates and measurements of SMBH
mass, BT10 found that the relationship between \ngc\ and \mbh\ was
even tighter than the $M$-$\sigma$ relation for those same galaxies.
BT10 also found that the mass of the central SMBH in the elliptical
and S0 galaxies they studied was roughly equal to the mass in GCs.
They concluded that the origin of the \ngc$-$\mbh\
correlation was obscure and that a larger sample of galaxies was
needed to further explore its causes and implications.

Harris \& Harris (2011; hereafter HH11) followed up on this previous
work
by expanding the sample of galaxies to more than twice that of BT10.
Starting with the list of galaxies for which \mbh\ had been
measured, they 
%
selected from the literature those with either published \ngc\
values or with information about the GC system that would allow them
to produce at least rough estimates of \ngc.
(BT10 had restricted their sample to galaxies included in the ACS
Virgo Cluster Survey paper on GC populations by Peng et al.\ 2008 or
the list of GC system properties compiled by Spitler et al.\ 2008.)
The end result was a sample of 33 elliptical, S0, and spiral galaxies
with both \mbh\ and \ngc\ values.  HH11 used the elliptical galaxies
in this sample to derive a relation between \ngc\ and \mbh\ and found
that it was consistent with the relation from BT10 within the errors.
They also examined how the correlation varied with the morphology of
the host galaxies.
HH11 reasoned that correlations between GC populations and SMBHs may
arise because of similarities in the conditions and epoch of formation
of these two types of objects.  They noted that both likely originate
in situations in which high gas densities are produced by energetic
collisions and mergers of gas clouds.  Such circumstances would be
expected to occur in the early universe (e.g., $z$ $\sim$10$-$15 or
higher), during the initial phase of galaxy assembly and formation.


Snyder, Hopkins, \& Hernquist (2011) drew on the data sets and results
of both BT10 and HH11 in order to further examine the possible causes
of the \ngc$-$\mbh\ correlation. They argued that the observed
correlation is a result of a link between SMBH mass and the binding
energy of the host galaxy bulge -- the so-called Black Hole
Fundamental Plane (BHFP; e.g., Hopkins et al.\ 2007) -- combined with
a link between \ngc\ and the stellar mass of galaxy bulges at fixed
velocity dispersion.  They examined in detail the residuals in the
\ngc$-$\mbh\ correlation as well as
correlations between \ngc\ and other galaxy properties.  Snyder et
al.\ showed that although the scatter in the \ngc$-$\mbh\ correlation
is small, this does not imply a ``special'' relationship betwen GCs
and SMBHs, but instead is due to both \mbh\ and \ngc\ correlating with
bulge mass. They point out that while the relationship betwen SMBH
mass and bulge potential is consistent with our general understanding
of galaxy mergers and physical processes like black hole accretion, feedback,
and gas cooling, the reasons why \ngc\ should be so tightly correlated
with bulge stellar mass are not entirely clear.

The measurements of \ngc\ used in these papers
are taken from heterogeneous data sets of varying quality.  Therefore
a key question that should be explored is what would happen to
the relation between \ngc\ and \mbh, as well as the relations between
\ngc\ and other fundamental galaxy properties, 
if the galaxy sample were 
constrained to include only galaxies with
the most well-determined GC system properties.
%
%
%
%
Indeed, BT10 conclude their paper by noting that, ``An important next
step is to expand the sample of galaxies having both reliable SMBH
masses and reliable GC populations.''  I am leading an ongoing
wide-field CCD imaging survey aimed at establishing the total numbers
and other global properties of the globular cluster systems of giant
galaxies
(Rhode \& Zepf 2001, 2003, 2004; Rhode et al.\ 2005, 2007, 2010;
Hargis et al.\ 2011; Young, Dowell, \& Rhode 2012, Hargis \& Rhode
2012).
Using the survey results 
to investigate 
the issues raised by BT10, HH11, and Snyder et al.\ (2011) 
is a natural application of the 
data.
Accordingly, I have compiled a sample of 20 galaxies with
measured GC system properties from observational studies that
meet specific criteria -- i.e., 
studies
in which the imaging data cover a fair fraction of the GC system and
in which quantities like the global color fraction are known.  To
date, we have observed $\sim$25 galaxies for the wide-field imaging
survey; eight of these have SMBH mass determinations in the
literature, so they make up eight of the 20 galaxies included in this
paper.
%
%
%
%
The remaining twelve galaxies included here are drawn from several
GC system studies in the literature.

Here I present the results of this investigation.
In Section 2, I discuss how \ngc\
and other GC system properties are measured in giant galaxies, what
issues give rise to uncertainties in those numbers, and the criteria I
applied to decide what data to include in the current study.
I then present the data set, including 
details about
how measurements were derived for each individual galaxy.
Section 3 describes the results of the study: I examine correlations
between \ngc\ and SMBH mass for the sample galaxies, investigate the
link between metal-poor and metal-rich GCs and SMBH mass, and look for
trends in the \ngc$-$\mbh\ correlation with galaxy morphology and
bulge type.  I also investigate the link between the numbers of GCs
and host galaxy properties.  In the last subsection of Section 3, I
calculate predicted SMBH masses for a set of galaxies from my
wide-field survey.  The final section of the paper gives a brief
summary of the main conclusions.


\section{The Data Set}
\label{section:data set}

\subsection{Deriving Global Properties of GC Systems}
\label{section:global properties}


Quantifying the properties of the GC systems of giant galaxies beyond
the Local Group is a challenging task.  Although it has been known for
decades that giant galaxies besides the Milky Way and M31 {\it host}
GCs -- e.g., authors like Baum (1955), Sandage (1961), and Dawe \&
Dickens (1976) speculated that the point-like objects surrounding
galaxies in the Virgo and Fornax Clusters were GCs -- it wasn't until
the early 1980s that systematic studies of GC populations in large
numbers of galaxies began in earnest (Harris 1991). GCs in galaxies
more than a few Mpc away
are unresolved in typical ($\sim$0.5$-$1.0\arcsec) ground-based
seeing; the median half-light radius of a Milky Way globular cluster
is 3 pc, which translates to $\sim$0.04\arcsec\ at the distance of the
Virgo Cluster (Ashman \& Zepf 1998).
The luminosities of GCs follow a roughly Gaussian distribution
referred to as the GC Luminosity Function (GCLF), which has a peak
magnitude $M_V^0$ $=$ $-$7.4$\pm$0.25 and a dispersion $\sigma$ $=$
1.4$\pm$0.2 mag (e.g., Whitmore et al.\ 1995, Kavelaars et al.\ 2000,
Barmby et al.\ 2001, Kundu \& Whitmore 2001, Jordan et al.\ 2007).
Therefore, studies of extragalactic GC systems aim to identify GCs as
a population of
point sources with optical magnitudes and colors like GCs arrayed in a
centrally-concentrated distribution around the host galaxy.  To derive
the total number of GCs in a galaxy's system, one first identifies and
counts the GCs (trying to distinguish them from other compact objects
like foreground stars or background galaxies) and then makes various
corrections to account for missing spatial coverage, the detection
limits of the images, and contamination from non-GCs, to arrive at a
final estimate of \ngc.

Early studies of extragalactic GC systems
employed photographic plates and source detection in one or two
filters to identify GCs and separate them from contaminating objects.
GCs can be found tens to hundreds of kiloparsecs from the center of
their host galaxy
(Harris 1991; Ashman \& Zepf 1998) and large-format
  photographic plates were a good match for
the extended nature of giant galaxy GC systems. On the other hand, the
low quantum efficiency of plates meant that only the brightest
portion
of the GC population could be detected. Photographic studies
also suffered from high levels of contamination from stars and
galaxies.
The result was that quantities like N$_{\rm GC}$ and color
distributions for the full system were often highly uncertain and
based on observations of only a small percentage of the total GC
population.  
A few examples arbitrarily selected from the review
article by
Harris (1991) serve to illustrate this.
Harris summarized the progress of extragalactic GC system studies and
compiled estimates of \ngc\ from the literature, listing both number
of clusters observed ($N_{obs}$) and the number of GCs in the system.
The latter quantity is computed by extrapolating $N_{obs}$ over all
magnitudes and radii.  For the Sab galaxy NGC~4569, $N_{obs}$ $=$
30$\pm$10 and the derived number of GCs
is 1000$\pm$400. For the S0 galaxy NGC~3607, $N_{obs}$
$=$ 50$\pm$35 and derived \ngc\
$=$ 800$\pm$600; for the elliptical NGC~3311, $N_{obs}$ $=$ 414$\pm$31
and 
the estimated \ngc\ is 41 times larger, at 17000$\pm$6000.

The increased availability of CCD detectors
in the late 1980s made detecting faint GCs in external galaxies more
efficient and made it easier to use color criteria (e.g., $B-V$) to
select GCs, thereby reducing contamination from Galactic stars and
background galaxies. Early CCDs had small formats, however, which
meant that large radial extrapolations were necessary to derive global
GC system properties. For instance, a careful imaging study by Lee,
Kim, \& Geisler (1998), performed using a 2048 x 2048-pixel CCD and
Washington $C$ and $T1$ filters, traced the GC system of the Virgo
elliptical NGC~4472 (M49) to 7\arcmin ($\sim$34~kpc).  Lee et
al.\ calculated both \ngc\ and specific frequency $S_N$ for NGC~4472's
GC system.  Specific frequency was introduced by \citet{hvdb81} and is
the number of GCs normalized by the host galaxy V-band luminosity:

\begin{equation}
{\rm{S_N \equiv {N_{GC}}10^{+0.4({M_V}+15)}}}
\label{equation:S_N}
\end{equation}

\noindent To calculate total number and $S_N$, Lee et al.\ integrated
their observed GC system radial distribution to 10\arcmin. Later
$BVR$ imaging with an 8192 x 8192-pixel mosaic CCD camera showed that
NGC~4472's GC system actually extends to
at least 23\arcmin, or \gapp110~kpc (Rhode \& Zepf 2001).  

Beginning in the early 1990s, the superior resolution and sensitivity
of the Hubble Space Telescope (HST)
cameras enabled much deeper imaging of extragalactic GC systems as
well as allowing faint background galaxies to be resolved and
therefore eliminated from GC studies.  In many ways these advances
revolutionized the study of GC populations beyond the Local Group,
allowing the GCLF, GC color distributions, and even GC sizes to be
studied in detail in many galaxies (see Brodie \& Strader 2006 and
references therein).  Even so, HST's small field-of-view (FOV) meant
that typically only a fraction of a galaxy's GC system was imaged,
making it difficult or impossible to quantify global quantities like
total number and color distribution for the system.  For example, the
HST ACS Virgo Cluster Survey (ACSVCS; C\^ot\'e et al. 2004) imaged 100
Virgo ellipticals and S0 galaxies to radii of 2.4\arcmin\ (12 kpc),
whereas early-type galaxy GC systems at Virgo Cluster distances often
extend an order of magnitude beyond that, to 10\arcmin$-$20\arcmin, or
$\sim$50$-$100~kpc (Rhode \& Zepf 2001, 2004).  To arrive at estimates
of \ngc\ and global $S_N$ for galaxies from the ACSVCS, Peng et
al.\ (2008) supplemented their survey data with parallel WFPC2 imaging
(that provided observations of the outer regions of the GC systems of
massive galaxies) and ground-based surface density profiles from
wide-field imaging studies in the literature.  Some of the \ngc\ and
$S_N$ values derived from the ACSVCS end up being quite similar to
previously-published values, while others are significantly different.
For example, the Harris (1991) literature compilation lists the number
of GCs
in the Virgo E6 galaxy NGC~4564 as \ngc\ $=$ 1200$\pm$400; Peng et
al.\ estimate \ngc\ $=$ 213$\pm$31 for this galaxy.  Harris (1991)
lists 
\ngc\ $=$ 2200$\pm$440 for the Virgo E5 galaxy NGC~4621 and Peng et
al.\ derive \ngc\ $=$ 803$\pm$305.  The Harris (1991) table lists
\ngc\ $=$ 3500$\pm$1200 for the Virgo E2 galaxy NGC~4365 and Peng et
al.\ measure a similar number, although with a much smaller error:
\ngc\ $=$ 3246$\pm$598.  The variation in these published values
serves
%
%
both to highlight the very real difficulty of measuring 
global GC system properties (especially \ngc\ and $S_N$) and to
motivate why it is important to critically examine
how the numbers were derived before using them to study other
properties of the parent galaxies.
%
%
%
%

In the late 1990s, large-format and mosaic CCD imagers became
available on many 4-meter-class telescopes, making it possible to
image much larger portions of giant galaxy GC systems outside the
Local Group in one or a few pointings. Because of the improved
efficiency of this approach --- both in terms of increased sensitivity
compared to photographic studies, and increased areal coverage
compared to small-format CCDs or HST imaging --- it also became more
common to image the GC populations in more than two
broadband filters (e.g., Rhode \& Zepf 2001, Tamura et al.\ 2006,
Spitler et al.\ 2008).  The observational goals of these types of
studies typically were to cover the majority of the spatial extent of
the GC systems, to do deep photometry in order to sample a significant
fraction of the GC luminosity function (GCLF), and to reduce
contamination from stars and galaxies by selecting GCs via their
magnitudes and colors in multiple filters.  The scientific objectives
were to accurately measure global values for \ngc, $S_N$, and the
distributions of GC colors (which should vary primarily due to the
metallicities of the clusters, with metal-poor GCs having blue
broadband colors and metal-rich GCs appearing red) as well as to
examine how these quantities change with galactocentric radius.  As
Brodie \& Strader (2006) note in their review article on extragalactic
GC systems, relatively few of these types of studies have been done to
date, but
the pattern so far is that the
\ngc\ and $S_N$ values they produce are often smaller than past
values and the errors on \ngc\ and $S_N$
are typically reduced by a factor of two or more (e.g., Rhode \& Zepf
2001, 2003, 2004, Spitler et al.\ 2008).
For example, Gomez \& Richtler (2004) used the Calar Alto 3.5-m
telescope and MOSCA CCD detector to image three pointings around the
Virgo galaxy NGC~4374.  They observed the GC system radial profile out
to 12$\arcmin$ ($\sim$60 kpc) and derived \ngc\ $=$ 1775$\pm$150 and
$S_N$ $=$ 1.6$\pm$0.3, whereas the previous value was \ngc\ $=$
3400$\pm$800 and $S_N$ $=$ 5.6$\pm$1.3 from a photographic study by
Hanes (1977).  

It is likely that a number of factors contribute to the 
smaller total numbers and specific frequencies
derived from modern wide-field CCD studies.  One reason is that
contamination from non-GCs is significantly reduced in modern studies
because GCs are being selected according to their magnitudes in two or
more filters and because good image resolution (\lapp1'') allows many
background galaxies to be resolved and discarded from GC lists.
Another contributing factor is that power laws are often used in
small-field imaging studies to fit the radial profile of a GC system.
The power law, of the form log~$\sigma_{\rm GC}$ $=$ $a0$ $+$
$a1$~log~$r$, where $r$ is projected radius and $\sigma_{\rm GC}$ is
the surface density of GCs, is then integrated over all projected
radii, 
or out to some assumed value for the radial extent of the
system.  We have found in our ongoing GC system survey that
de~Vaucouleurs law profiles of the form log~$\sigma_{\rm GC}$ $=$ $a0$
$+$ $a1$~$r^{1/4}$ are often a better fit for GC system profiles out
to large radius, and these drop off more quickly in the outer regions
than a power law profile would (e.g., Rhode \& Zepf 2004).
Furthermore, we integrate the profiles only out to the radius where
the GC system surface density falls off to zero within the errors,
rather than integrating over all radii or to some arbitrary
value. Both of these issues likely contribute to reduced \ngc\ and
$S_N$ values and uncertainties derived from wide-field studies that
trace the full extent of the GC system.  In some cases the newer
studies have deeper imaging that enables a better determination of
what fraction of the GCLF has been imaged, which in turn yields a more
accurate correction for magnitude incompleteness.  Finally, in a few
cases -- for example, a 2003 study of the GC system of the Fornax
galaxy NGC~1399 by Dirsch et al. --- a smaller or more accurate
GC specific frequency may be derived simply because the total galaxy
magnitude was revised based on improved wide-field CCD surface
photometry.

Finally, it is worthwhile to briefly note here the effect that errors
in measured galaxy distances have on determinations of \ngc\ and $S_N$
for extragalactic GC systems.  Authors of GC system studies typically
adopt a distance to the host galaxy from the literature.  This
distance is then folded into the calculations of distance-dependent
quantities like the total galaxy magnitude and the fractional coverage
of the theoretical GCLF.  The latter quantity is usually determined by
(1) assuming a value for the peak absolute magnitude and dispersion of
a Gaussian GCLF and (2) fitting the observed luminosity function of GC
candidates to the Gaussian by varying the normalization and then
calculating how much of the area under this theoretical curve has been
covered by the data. The absolute peak magnitude and/or dispersion of
the theoretical GCLF are sometimes also varied by a few tenths of a
magnitude.  Typically \ngc\ and $S_N$ determinations in the literature
include errors on the counts of GC candidates and contaminants and
perhaps some modest uncertainty associated with the GCLF fitting
process,
but do not explicitly include the uncertainties in the galaxy distance
(see, e.g., Harris 1991, Ashman \& Zepf 1998, Spitler et al.\ 2008 and
other compilations of GC system properties).
%
If the distance to a galaxy has been underestimated and the {\it true}
distance is greater than assumed, the calculated fractional GCLF
coverage will be larger than it should be and the final \ngc\ will be
an {\it underestimate} of the true value. (That is, if the distance
were corrected to its larger value, the final \ngc\ would increase.)
%
Because both the galaxy magnitude and \ngc\ contribute to $S_N$,
distance errors have a different net effect on specific frequency
estimates.
%
A galaxy that in reality is farther away than the adopted distance
will have an underestimated \ngc, but its intrinsic luminosity will
likewise be underestimated. Changes in \ngc\ and $M_V$ in
Equation~\ref{equation:S_N} will counteract each other, so an increase
in distance to the galaxy can produce a larger \ngc\ but a smaller GC
specific frequency (see an example of this in our study of the GC
system of NGC~7814; Rhode \& Zepf 2003).

\subsection{Constructing the Sample} 

\subsubsection{Criteria Used to Select Galaxies}
\label{section:criteria}

Given the issues described in the previous section, my objective
was to compile
a sample of giant galaxies with well-determined measurements of
\ngc\ and GC specific frequency, drawing from my own survey and from the
literature.  I also decided to restrict the sample to those galaxies
for which the global GC color distribution (i.e., the number of GCs
versus broadband color, which is an indicator of metallicity) is
known. Model scenarios for the formation and evolution of GC systems
(e.g., Ashman \& Zepf 1992, Cote at el.\ 1998, Forbes et al.\ 1997,
Beasley et al.\ 2002, Muratov \& Gnedin 2010) typically make
predictions for the ratio of blue (metal-poor) and red (metal-rich)
GCs and how that ratio should change with radius in different types of
galaxies.  Because the metal-rich GC population
has been shown to be more centrally-concentrated than the metal-poor
population in some galaxies (e.g., Lee et al.\ 1998, Rhode \& Zepf
2004), color distributions measured in only the central portions
of GC systems may not accurately represent the global color
distribution; thus wide-field coverage can also be important for
measuring the true {\it global} fractions of blue and red GCs in a galaxy.

Rhode, Zepf, \& Santos (2005) examined how the specific frequencies of
blue, metal-poor GCs varied with host galaxy stellar mass for a sample
of giant spiral, S0, and elliptical galaxies.  We formulated a set of
criteria that we believed would
limit the sample to well-determined measurements.  I have adopted the
same standards here. The GC system studies included in the current
sample must meet the following criteria:
(1) at least 50\% of the estimated radial extent of the GC system must
have been observed; (2) imaging data must have been acquired in at
least two filters, so that contamination from non-GCs can be reduced
and the GC color distribution can be quantified; (3) an estimate of
\ngc\ is given or can be derived in a straightforward way from the
published data, and (4) the 1-$\sigma$ error on \ngc\ or $S_N$ must be
\lapp40\%.  Since my aim for the current study is to use the galaxy
sample to investigate the connection between supermassive black holes
and GC populations, the galaxies in the sample must also have a
measurement of SMBH mass in the literature.  I used the lists of
galaxies in BT10 and HH11 for initial guidance regarding which
galaxies had SMBH mass measurements, but also looked through the
literature for additional measurements that may not have appeared in
either of those papers.
The papers from which I drew most of the SMBH mass measurements are
the compilations of Gultekin et al.\ (2009) and Graham (2008) and
references therein.  I also examined the updated compilation of SMBH
masses in Graham et al.\ (2011), but found no additional \mbh\ values
that would supplement the numbers taken from the Gultekin et
al.\ (2009) and Graham (2008) papers.

The final result of my search is a list of 20 galaxies that meet the
stated criteria for GC system observations and also have published
SMBH mass measurements.  The data for these galaxies and the
references from which the quantities are derived are given in
Table~\ref{table:master}.  The table lists, in this order: galaxy name
and morphological type; SMBH mass and associated reference(s);
velocity dispersion (for the galaxy or bulge, as appropriate); galaxy
absolute $V$ magnitude; the galaxy distance (in Mpc) that was assumed
for the GC system values; number of GCs; GC specific frequencies;
fraction of blue GCs in the system; and the reference for the GC
system properties. (Note that estimated uncertainties on the fraction
of blue or red GCs in the galaxies in this sample are typically a few
to $<$10\%; Rhode \& Zepf 2001, 2004, Peng et al.\ 2006).  The GC
specific frequency $S_N$ in column 9 of Table~\ref{table:master} is as
defined in Equation~\ref{equation:S_N}.  Another type of specific
frequency, $T$, is given in column 10. $T$ was suggested by Zepf \&
Ashman (1993) and is defined as:

\begin{equation}
{\rm{T \equiv \frac{N_{GC}}{M_G/10^9\ {\rm M_{\sun}}}}}
\label{equation:T}
\end{equation}

\noindent where \ngc\ is the number of GCs and $M_G$ is the stellar
mass of the host galaxy. Zepf \& Ashman (1993) point out that $S_N$ is
affected by variations in $V$-band stellar mass-to-light ratios for
galaxies of different morphological types and stellar populations, so
$T$ can be useful when one is comparing GC specific frequencies for
galaxies over a range of morphologies.
I have adopted the mass-to-light ratios used by Zepf \& Ashman when
they defined $T$: $M/L_V$~$=$~10 for ellipticals, $M/L_V$~$=$~7.6 for
S0 galaxies, and $M/L_V$ from 6.1 to 4.0 for spiral galaxies, with
$M/L_V$ decreasing with later Hubble type.

\subsubsection{Comments on Individual Galaxies}

\noindent {\it Galaxies from Our Wide-Field GC System Survey.---}  The GC
system values for eight of the galaxies in Table~\ref{table:master}
come from the wide-field GC system survey that I have been leading 
(Rhode \& Zepf 2001, 2003, 2004; Rhode et al.\ 2005, 2007, 2010;
Hargis et al. 2011; Young et al.\ 2012, Hargis \& Rhode 2012).
We image spiral, S0, and elliptical galaxies at distances of
$\sim$10$-$25~Mpc with large-format and mosaic CCD imagers. The FOVs
of the cameras we use are $\sim$10$-$36\arcm\ on a side, which
translates to $\sim$30$-$200~kpc at these distances; this is
sufficient to observe the full radial extent of the GC systems of the
target galaxies, which typically range from $\sim$ 10$-$100~kpc (e.g.,
Rhode et al.\ 2010). We select the GC candidates around each galaxy
via three-color photometry to both reduce contamination from non-GCs
and to allow us to investigate the color distributions and color
gradients of the GC systems. The resolution of our images is
\lapp1\arcs\ so we can eliminate most background galaxies from the GC
candidate lists.  This careful GC candidate selection, and the fact
that we can usually trace the spatial distributions of the GC systems
over their full radial extent, means that the errors on our derived
total numbers and specific frequencies are significantly smaller than
those from past studies, and meet the criteria listed in
Section~\ref{section:criteria}.

The GC system properties of NGC~1023, NGC~7332, NGC~7339, and NGC~7457
were derived from $BVR$ imaging acquired with the 4096 x 4096-pixel
Minimosaic camera on the WIYN 3.5-m telescope.\footnote{The WIYN
  Observatory is a joint facility of the University of Wisconsin,
  Indiana University, Yale University, and the National Optical
  Astronomy Observatory.} The GC populations of NGC~3379, NGC~3384,
NGC~4472, and NGC~5813 were observed in $BVR$ with the 8192 x
8192-pixel Mosaic imager on the Mayall 4-m telescope at Kitt Peak
National Observatory.

\medskip
\noindent {\it Galaxies from the ACS Virgo Cluster Survey.---}  The GC
system properties of NGC~4350, NGC~4459, NGC~4473, and NGC~4564 come
from the ACSVCS survey, which was mentioned briefly in
Section~\ref{section:global properties}. Peng et al.\ (2008) used the
ACSVCS data to derive total numbers and specific frequencies for 100
early-type galaxies in the Virgo Cluster.  The ACS FOV is 202\arcs\ x
202\arcs, so the radial coverage of the GC system for a galaxy
centered in the field is 143\arcs, which corresponds to 11.5~kpc at
16.5~Mpc, their assumed distance to Virgo.
In order for a galaxy
to be included in
Table~\ref{table:master}, at least 50\% of the radial extent of the GC
system must have been observed.  Therefore the relevant question is:
given the areal coverage of the Peng et al.\ study, for which of the
100 ACSVCS galaxies has this criterion been satisfied?

In Rhode et al.\ (2010), we used the results from our wide-field GC
system survey to derive a relationship between the stellar mass of a
galaxy and the radial extent of the galaxy's GC system.  Taking that
relation (Equation 1 in Rhode et al.\ 2010), I calculated the stellar
mass of a galaxy for which the ACSVCS imaging would encompass 50\% of
the radial extent of the GC system.  Galaxies with $log(M/M\odot)$
\lapp11.3 fall into this category. The giant galaxies in the ACSVCS
are all elliptical or S0 galaxies, so I converted this mass value to
an absolute
$V$-band magnitude by assuming mass-to-light ratios for E and S0
galaxies from Zepf \& Ashman (1993).
According to this calculation, the ACSVCS imaging should be sufficient
to meet the 50\% radial coverage criteria for ellipticals 
with $M_V$ $\geq$ $-$20.9 and S0s with $M_V$ $\geq$ $-$21.2.

I searched the list of ACSVCS galaxies 
to find E/S0 galaxies that meet these magnitude criteria, have
measured SMBH masses, and were not already included in my sample.  The
result is a list of six galaxies out of the 100 ACSVCS targets.  Two
of these galaxies, NGC~4486A and NGC~4486B, are ellipticals that are
companions to the giant elliptical NGC~4486 (M87).  Peng et
al.\ (2008) estimate the total number of GCs in NGC~4486A and
NGC~4486B is \ngc\ $=$ 11$\pm$12 and \ngc\ $=$ 4$\pm$11,
respectively. The galaxies therefore fail one of the stated criteria,
that the error on \ngc\ must be \lapp40\%. Four ACSVCS galaxies ---
the two ellipticals NGC~4473 and NGC~4564 and the two S0 galaxies
NGC~4459 and NGC~4473 --- meet all of the requirements and are thus
included in Table~\ref{table:master}.
%
The \ngc, $M_V$ values, and blue GC fractions in
Table~\ref{table:master} are taken directly from Peng et al.\ (2008).
I combined \ngc\ and $M_V$ with my assumed $M/L_V$ values to calculate
$S_N$ and $T$ for the table.

\medskip
\noindent {\it Galaxies from the Gemini Study of Faifer et al.---} Two
of the galaxies listed in Table~\ref{table:master} come from a study
by Faifer et al.\ (2011), who used the GMOS camera on both the Gemini
North and South telescopes to image the GC systems of early-type
galaxies in the g', r', and i' broadband filters. They estimate
\ngc\ for two galaxies that have published SMBH masses: NGC~3115 and
NGC~4649.  Based on the relation between galaxy stellar mass and GC
system extent from Rhode et al.\ (2010), the GC system of the S0
galaxy NGC~3115 should extend to $\sim$26~kpc from the galaxy center,
while the elliptical galaxy NGC~4649's GC system should go out to a
radius of $\sim$73~kpc.  Faifer et al.\ trace the radial distribution
of the GC population of NGC~3115 to $\sim$5\arcm\ (14~kpc) and the GCs
in NGC~4649 to $\sim$8\arcm\ (40~kpc), so the radial coverage
requirement appears to have been met.
The galaxies were imaged in the g', r', and i' broadband filters and
Faifer et al.\ provide values for the fraction of blue GCs in the
system.
%

Faifer et al.\ fitted both power laws and deVaucouleurs profiles to
the radial distributions of the GC systems of the target galaxies, and
integrated the profiles to calculate \ngc\ and $S_N$ for two radial
limits, 50~kpc and 100~kpc, presumably because they were uncertain
about how far out the GC systems extended.  To calculate \ngc\
for NGC~3115, I integrated the best-fitting deVaucouleurs law profile
from Faifer et al.\ out to 26~kpc (9.3\arcm\ for a distance to the
galaxy of 9.7~Mpc).  I combined this with their assumed absolute
magnitude for NGC~3115 ($M_V$ $=$ $-$21.13) and $M/L_V$ $=$ 7.6 to
calculate $S_N$ and $T$.  I went through similar steps for NGC~4649,
this time integrating the best-fitting deVaucouleurs GC system profile
to 73~kpc (15.0\arcm\ for the assumed distance of 16.8~Mpc) and using
$M_V$ $=$ $-$22.38 and $M/L_V$ $=$ 10 to calculate \ngc, $S_N$, and
$T$.  The final values for these two galaxies appear in
Table~\ref{table:master}.

\medskip
\noindent {\it Galaxies from Other Studies in the Literature.---}  
The GC system properties of the remaining six galaxies in the sample
of 20 were drawn directly from various studies in the literature that
satisfy the selection criteria
outlined in Section~\ref{section:criteria}. These galaxies also have
published SMBH masses. The references from which these quantities are
taken are listed in the table.  Note that the \ngc\ and specific
frequency values for the GC system of the Milky Way are taken from
Ashman \& Zepf (1998), who based their estimates on mulitiple studies
of the Galactic GC system.  The \ngc\ and specific frequencies for
M31's GC population are calculated by combining the results given in
Ashman \& Zepf (1998) and Barmby et al.\ (2000) (for the total number
estimates and blue fraction) with Perrett et al.\ (2002) (for the
metal-poor/blue GC fraction).
%
%

\subsubsection{Rescaling the SMBH Values to Match the Distances Assumed for the GC System Studies}

When I assembled the data for Table~\ref{table:master}, I noticed that
the distances to the galaxies that were assumed for the GC system
studies were often different --- sometimes by as much as $\sim$10\%
--- from the galaxy distances assumed in the studies from which the
\mbh\ estimates were drawn. To make sure that any observed spread in
the \ngc-\mbh\ points was not due to differences in the adopted
distance, I have rescaled the SMBH masses to the GC system distances
(which appear in column 7 of Table~\ref{table:master}).
I followed the general approach of Gultekin et al.\ (2009) and assumed
that the \mbh\ values are proportional to the distance, and calculated
rescaled \mbh\ numbers by multiplying them by the ratio of the GC
distance to the original assumed SMBH distance.
The SMBH values listed in column 3 of Table~\ref{table:master} are
these rescaled values.  It is worth noting that the scale factors
applied to the SMBH masses ranged from 0.88$-$1.08 (with most in the
range 0.95$-$0.99) and the resultant changes to the \mbh\ values used
here were small and well within the stated errors on the \mbh\ values
given in the literature.

\section{Results}


\subsection{Relation between \ngc\ and \mbh}
\protect\label{section:ngc mbh}

The main result discovered by BT10 and investigated by HH11 and Snyder
et al.\ (2011) is the correlation between the log of \ngc\ and the log
of the SMBH mass for early-type galaxies. Figure~\ref{fig:ngc smbh}
shows this correlation for my sample of giant E, S0, and spiral
galaxies.  Three of the galaxies in the sample --- NGC~1399, NGC~3379,
and NGC~5128 --- are plotted twice.  This is because they each have
two SMBH mass determinations that significantly differ from one
another but are deemed individually reliable in the compilation of
Gultekin et al.\ (2009).  Following that paper and BT10, I show both
SMBH masses in the plots and have used both estimates in my fits,
assigning each of the points half-weight. Red filled circles denote
galaxies with classical bulges and green open triangles mark galaxies
with bulges that are sometimes classified in the literature as
possessing a pseudobulge; this issue will be discussed in more detail
in Section~\ref{section:pseudo classical}.

\begin{figure}
\includegraphics[width=3.6in]{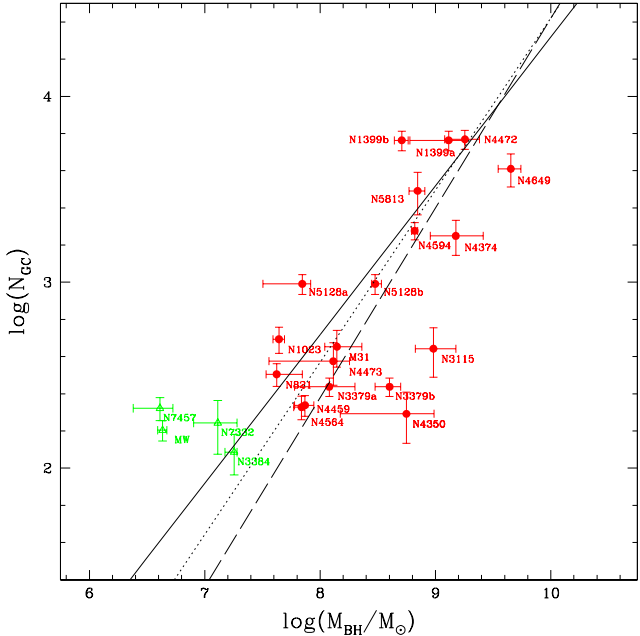}
\caption{\normalsize The log of the number of GCs versus the log of
  the SMBH mass for the 20 giant galaxies included in the final sample
  and listed in Table~\ref{table:master}. The red filled circles mark
  galaxies with classical bulges and the green open triangles denote
  galaxies that may have pseudobulges, according to the
  literature. NGC~1399, NGC~3379, and NGC~5128 each have two values of
  SMBH mass that are given half-weight in the fitting process (see
  Table~\ref{table:master} for more information). The solid line is
  the best-fit line for the 18 E/S0 galaxies in the sample
  (Equation~\ref{equation:ES0 bf}).  The dotted line is the best-fit
  line to the E/S0 galaxies from BT10 and the dashed line is the
  best-fit relation for E galaxies from HH11.}  \protect\label{fig:ngc
  smbh}
\end{figure}


So that I can compare my results directly to those of the previous
studies on this topic, I have used the same statistical analysis
methods employed by BT10 and HH11. Specifically, I have
fitted lines to the data included in Figure~\ref{fig:ngc smbh} using
the $\chi^2$-minimization method described in Tremaine et al.\ (2002)
and Press et al.\ (1992).  The method, called {\tt fitexy} in Press et
al.\ (1992), computes a best-fit linear relation of the form:

\begin{equation}
{\rm y = \alpha + \beta x} ,
\label{equation:line}
\end{equation}

\noindent while taking into account the errors in both $x$ and $y$.
The quantity 

\begin{equation}
\rm{\chi^2 = \sum_{i=1}^{N}\frac{{(y_i - \alpha - \beta x_i)}^2}{\epsilon_{yi}^2 + \beta^2\epsilon_{xi}^2}}
\label{equation:chisq}
\end{equation}

\noindent is minimized in the fitting process; here, $\epsilon_{xi}$
and $\epsilon_{yi}$ are the errors on $x$ and $y$, respectively.  The
errors are assumed to be symmetric,
so I have calculated error bars on logarithmic quantities by averaging
the high and low error bars; i.e., the error on each quantity is:

\begin{equation}
\rm{{\sigma_{log~i} = {1/2}[log(i + \sigma_i) - log(i - \sigma_i)]}},
\end{equation}

\noindent where $i$ in this case is either \ngc\ or \mbh.

BT10 included thirteen elliptical and S0 galaxies in their
\ngc-\mbh\ fit; their best-fit relation is plotted as a dotted line in
Figure~\ref{fig:ngc smbh} and is given in their paper in the form:

\begin{equation}
\rm{{log\frac{M_{\rm SMBH}}{M_{\odot}} = (8.14\pm0.04) + (1.08\pm0.04) log\frac{N_{GC}}{500}}}
\protect\label{equation:BT10 ES0 bf}
\end{equation}

\noindent HH11 fitted the \ngc-\mbh\ correlation using 18 ellipticals
from their sample of 33 galaxies and termed this 
their ``baseline relation''. It is plotted as a dashed line in
Figure~\ref{fig:ngc smbh} and can be written (following the format of
the best-fit line from BT10) as:

\begin{equation}
\rm{{log\frac{M_{\rm SMBH}}{M_{\odot}} = (8.30\pm0.29) + (0.98\pm0.10) log\frac{N_{GC}}{500}}}
\protect\label{equation:HH11 E bf}
\end{equation}


\noindent The galaxies with log(\mbh/$M_\odot$) greater than $\sim$7.5 in
Figure~\ref{fig:ngc smbh} generally do seem to follow the linear
relations of BT10 and HH11, but with a few galaxies (NGC~4649,
NGC~3115, and NGC~4350) deviating from the lines.  The group of four
galaxies with the lowest SMBH masses (NGC~7457, NGC7332, the Milky
Way, and NGC3384) also lie somewhat off the BT10 and HH11 relations.
If I follow BT10 and include all 18 E/S0 galaxies in my fit, I derive the
following best-fit relation:

\begin{equation}
\rm{{log\frac{M_{\rm SMBH}}{M_{\odot}} = (8.04\pm0.03) + (1.22\pm0.06)
  log\frac{N_{GC}}{500}}}
\protect\label{equation:ES0 bf}
\end{equation}

\noindent This relationship is plotted as a solid line in
Figure~\ref{fig:ngc smbh}.  It has a steeper slope than the relations
from both BT10 and HH11, although the slope and intercept differ from
the previous values by less than 3$\sigma$, so
are nevertheless consistent within the errors.

To assess the quality of the fits in their paper, BT10 calculated the
$\chi^2$ per degree of freedom, 
which is often called the ``reduced $\chi^2$''.  Normally when one
fits a line, the number of degrees of freedom is $N-2$, where $N$ is
the number of data points included in the fit.  BT10 used $N-4$ to
normalize their $\chi^2$ values
because three of the points they used are related (the three galaxies
that each have two SMBH mass determinations in
Table~\ref{table:master}) and the inclusion of these points likely
reduces the true number of degrees of freedom (Tremaine 2011, private
communication). For their sample of 13 E/S0 galaxies,
BT10's best-fit \ngc-\mbh\ line yields a reduced $\chi^2$ $=$
$\chi^2/(N-4)$ $=$ 6.6. For my sample of 18 E/S0 galaxies (including
the same three galaxies with multiple SMBH measurements)
my best-fit relation 
yields $\chi^2/(N-4)$ $=$ 10.4 and $\chi^2/(N-2)$ $=$ 9.3.
Therefore, although I have limited my sample to galaxies with
relatively well-determined \ngc\ values, the result is a
\ngc-\mbh\ fit of lower quality (as measured by the reduced $\chi^2$)
than the fit produced by BT10 for their galaxy sample.  On the other
hand, the errors on my best-fit slope and intercept in
Equation~\ref{equation:ES0 bf} are smaller than the errors on those
quantities in the HH11 fit and comparable to the errors in the BT10
fit.

One way that previous authors have used to gauge the strength or
quality of the correlation, as well as to account for the possibility
that there is an intrinsic dispersion in the \ngc-\mbh\ relation,
is to add an additional dispersion term to one of the measurement
errors and repeat the fitting process (cf.\ Tremaine et al.\ 2002,
BT10, Snyder et al.\ 2011).  The assumptions here are that the
measurement errors are accurate and that there is some underlying,
real galaxy-to-galaxy scatter present in the galaxy properties.
With this type of approach, the amount of additional scatter that
produces a reduced $\chi^2$ $\sim$ 1 is inferred to be the true
intrinsic dispersion of the quantity.  To implement this, one replaces
the quantity $\epsilon_{yi}^2$ in Equation~\ref{equation:chisq} with
$(\epsilon_{yi}^2 + \epsilon_0^2)^{1/2}$ and performs the fitting
procedure, adjusting $\epsilon_0$ until the value of the reduced
$\chi^2$ is 1.
Smaller values of $\epsilon_0$ are then interpreted as indicating a
tighter correlation between the two quantities in the fit. 

For my sample of 18 E/S0 galaxies, adding an intrinsic scatter of
$\epsilon_0$ $\sim$ 0.45 dex in quadrature to either log~\ngc\ or
log~\mbh\ produces a fit with reduced $\chi^2$ $\sim$ 1. The slope and
intercept of the resultant fit are (using the notation from
Equation~\ref{equation:line}) $\beta$ $=$ 1.11$\pm$0.19 and $\alpha$
$=$ 8.10$\pm$0.11, which are the same within the errors as the slope
and intercept derived with no additional intrinsic scatter
(Equation~\ref{equation:ES0 bf}).  For completeness, I should note
that another approach sometimes seen in the literature is to simply
add an intrinsic scatter term {\it directly} to the measurement
errors, i.e., $\epsilon_{yi}$ $=$ $\epsilon_{yi}$ $+$ $\epsilon_0$ and
then perform the fitting process.  In this case, $\epsilon_0$ $\sim$
0.3~dex (which corresponds to a linear factor of $\sim$2) needs to be
added to the errors in order to produce reduced $\chi^2$ of unity for
the E/S0 galaxy sample.

When I include all the galaxies in the sample (20 ellipticals, S0s,
and spiral galaxies) and calculate the best-fit relation between
\ngc\ and \mbh\, the result is:

\begin{equation}
\rm{{log\frac{M_{\rm SMBH}}{M_{\odot}} = (7.91\pm0.03) + (1.52\pm0.06)
  log\frac{N_{GC}}{500}}}
\protect\label{equation:all bf}
\end{equation}

\noindent Including the two spiral galaxies in the fit steepens the
slope of the relation not because of the data point for M31, which
lies very close to the best-fitting relation for the E/S0 galaxies,
but because of the data point for the Milky Way, which has
comparatively small error bars and deviates significantly from the
best-fitting E/S0 relation.  The quality of this fit (as measured by
reduced $\chi^2$)
is degraded compared to the fit that includes only E/S0 galaxies;
here, $\chi^2$/$(N-4)$ $=$ 12.4 and $\chi^2$/$(N-2)$ $=$ 11.2. The
scatter as measured by the $\epsilon_0$ parameter is also larger: one
must add $\epsilon_0$ $=$ 0.50 in quadrature to the errors in order to
produce a reduced $\chi^2$ value of $\sim$1.  Note that the results
(slopes, intercepts, $\chi^2$ values, etc.) of the
\ngc$-$\mbh\ fitting process described in this section and in
later sections of the paper are summarized in
Table~\ref{table:fitting
  results}. The table lists: the sample used in the fitting process,
the number of points in the sample, the intercept and slope of the
best-fitting relation, the reduced $\chi^2$
values, and the value of the intrinsic scatter ($\epsilon_0$).

\subsection{$M$-$\sigma$ Relation for the Galaxy Sample}

\begin{figure}
\includegraphics[width=3.5in]{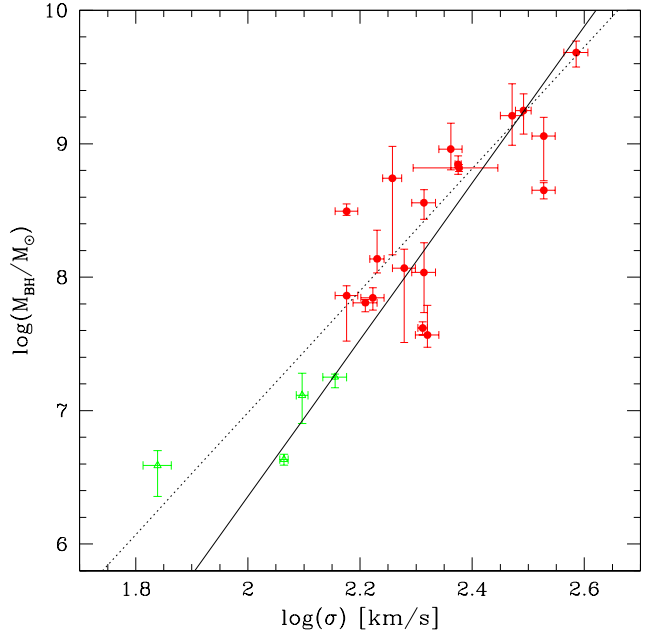}
\caption{\normalsize The log of the SMBH mass versus the log of the
  velocity dispersion of the galaxy (for early-type galaxies) or the
  bulge (for spiral galaxies) for the 20 giant galaxies in
  Table~\ref{table:master}. Red filled circles denote galaxies with
  classical bulges and green open triangles represent galaxies
  classified in the literature as having pseudobulges.  Three galaxies
  are plotted twice and given half-weight in the fitting process (see
  Table~\ref{table:master}). The solid line is the best-fitting line
  for all 20 galaxies (Equation~\ref{equation:m-sigma all}) and the
  dotted line is the best-fit $M$-$\sigma$ relation from BT10.  }
\protect\label{fig:m sigma}
\end{figure}

Another useful or relevant comparison comes from determining, for my
galaxy sample, whether a tighter correlation exists for \ngc-\mbh\ or
for \mbh-$\sigma$. BT10 found that the former produced a
higher-quality fit and therefore a stronger correlation.
Figure~\ref{fig:m sigma} shows the \mbh-$\sigma$ relation for my
20-galaxy sample. As in Figure~\ref{fig:ngc smbh}, three galaxies are
plotted twice and given half-weight in the fits.  Galaxies with
classical bulges are again represented by red filled circles and
galaxies with pseudobulges are plotted with green open triangles; see
Section~\ref{section:pseudo classical} for more discussion.

I used the {\tt fitexy} method to fit a line to the 20 galaxies in
Figure~\ref{fig:m sigma}.  The result is shown as a solid line in the
figure and can be written:

\begin{equation}
{\rm{log\frac{M_{\rm SMBH}}{M_{\odot}} = (8.12\pm0.03) + (5.87\pm0.20)
  log\frac{\sigma}{200~km~s^{-1}}}}
\protect\label{equation:m-sigma all}
\end{equation}

\noindent The fit yields reduced $\chi^2$ values of $\chi^2/(N-4)$ $=$
13.0 and $\chi^2/(N-2)$ $=$ 11.8. When I limit the galaxy sample to
only the 18 E/S0 galaxies (i.e., excluding the Milky Way and M31), the
best-fit line becomes:

\begin{equation}
{\rm{log\frac{M_{\rm SMBH}}{M_{\odot}} = (8.15\pm0.03) + (5.34\pm0.27)
  log\frac{\sigma}{200~km~s^{-1}}}}
\protect\label{equation:m-sigma E/S0}
\end{equation}
 
\noindent Both the error on the intercept and the reduced $\chi^2$
values become slightly larger in this case; here, $\chi^2/(N-4)$ $=$
13.8 and $\chi^2/(N-2)$ $=$ 12.4. For both the full galaxy sample and
the E/S0 sample, adding an additional scatter of $\epsilon_0$ $\sim$
0.4~dex in quadrature to the errors on the log of the SMBH masses
brings the reduced $\chi^2$ values to $\sim$1.
 
The corresponding best-fit $M$-$\sigma$ relation found by BT10 for
their sample of 13 E/S0 galaxies is plotted as a dotted line in
Figure~\ref{fig:m sigma} and is:

\begin{equation}
{\rm{log\frac{M_{\rm SMBH}}{M_{\odot}} = (8.36\pm0.04) + (4.57\pm0.25)
  log\frac{\sigma}{200~km~s^{-1}}}}
\protect\label{equation:m-sigma BT10}
\end{equation}
 
\noindent BT10 found that both the reduced $\chi^2$ value and the
amount of scatter in their best-fitting \ngc-\mbh\ relation were
smaller than those of their best-fitting $M$-$\sigma$ relation; thus,
they found that \ngc\ is a better predictor of SMBH mass than is
velocity dispersion $\sigma$. I find mixed results when I compare the
quality of the \ngc-\mbh\ and $M$-$\sigma$ correlations for my galaxy
sample.
%
%
As in BT10, the reduced $\chi^2$ values of my initial fits are
slightly smaller for the \ngc-\mbh\ correlation than for the
$M$-$\sigma$ relation.  The errors on the coefficients of my best-fit
lines are likewise smaller for the \ngc-\mbh\ relation than for the
$M$-$\sigma$ relation.  However, roughly the same amount of scatter
($\epsilon_0$ $\sim$ 0.4 dex) needs to be added in quadrature to the
errors on the data points in both the \ngc-\mbh\ and $M$-$\sigma$ fits
in order to bring the reduced $\chi^2$ values to unity.

Finally I should note for the sake of completeness that the slope of
the SMBH mass $-$ velocity dispersion relation derived for my sample
of E/S0 galaxies (5.34$\pm$0.27) is consistent within $<$3 times the
errors with the corresponding relation from BT10, as well as with the
derived slopes from other recent giant galaxy studies.
For example, Ferrarese \& Ford (2005) find the slope of the
$M$-$\sigma$ relation to be 4.86$\pm$0.43, Tremaine et al.\ (2002)
derive 4.02$\pm$0.32,
and Gultekin et al.\ (2009) calculate a slope of 3.96$\pm$0.42 for
ellipticals and 4.24$\pm$0.41 for both early- and late-types.  This is
not surprising, of course, since the SMBH mass and velocity data I
used were largely drawn from the same published measurements used by
these authors (although I have rescaled the SMBH masses to the
distances assumed for the GC system studies).

\subsection{Are Metal-Poor or Metal-Rich GCs Better at Tracing SMBH Mass?}

As I mentioned briefly in Section~\ref{section:criteria}, the 
broadband colors of GCs can be an indicator of their metallicities.
%
It is generally true that for old (\gapp2$-$3~Gyr) stellar
populations, optical colors primarily trace metallicity, with blue
colors corresponding to metal-poor populations and red to
metal-rich. This is caused by a combination of factors, such as line
blanketing due to the presence of metals in stellar atmospheres and
metal-rich stars having lower equilibrium temperatures on the main
sequence and in later evolutionary stages
(e.g., Schwarzschild et al.\ 1955, Sweigart \& Gross 1978).  The GC
systems of many giant galaxies have color distributions with more than
one peak (e.g., Zepf \& Ashman 1993, Kundu \& Whitmore 2001, Peng et
al.\ 2006), so these multiple peaks seem to imply the presence of GC
populations with different metallicities, presumably formed in
different episodes. Spectroscopic metallicities, kinematic studies,
and near-IR colors of GCs confirm this interpretation (e.g., Kundu \&
Zepf 2007, Strader et al.\ 2007, Beasley et al.\ 2008, Alves-Brito et
al.\ 2011).  The fact that the Milky Way and M31 also have two
populations of GCs with different metallicities and kinematics (Zinn
1985, Perrett et al.\ 2002) seems to suggest that multiple GC
populations may be a common characteristic of giant galaxies.

%
The overarching goal of many of the studies of extragalactic GC
systems in recent years has been to understand the range of giant
galaxy GC system properties --- including the presence of multiple
peaks in GC color/metallicity distributions --- 
within the larger context of galaxy formation and evolution (see the
review on this subject by Brodie \& Strader 2006).  Over time a
general picture has developed that incorporates aspects of a number of
different scenarios for GC and galaxy formation (e.g., Ashman \& Zepf
1992, Forbes et al.\ 1997, Cote et al.\ 1998, Beasley et al.\ 2002,
Santos 2003, Rhode, Zepf, \& Santos 2005, Muratov \& Gnedin 2010).
The precise details are not yet well-established,
but the 
data seem to point toward a picture (that Brodie \& Strader 2006 term
the``synthesis scenario'') in which metal-poor GCs were formed when
low-mass, gas-rich protogalactic fragments collided and merged at high
redshift ($z$$\sim$10$-$15) and metal-rich GCs were formed during
dissipational mergers at later times.  In this picture the blue,
metal-poor GC populations in giant galaxies are therefore fossils left
over from the first epoch of GC and galaxy assembly. The effect of
cosmological ``biasing'' may also be an important piece of the puzzle;
massive galaxy halos located in dense environments are expected to
have a larger proportion of their mass accumulated by a given redshift
than lower-mass galaxies in lower-density environments (e.g., Sheth
2003). 

In some models, this initial phase of GC formation is truncated or
suppressed by some mechanism around $z$$\sim$10 for a brief period of
time.
Santos (2003) suggests reionization would serve as such a truncation
mechanism, and would suppress both GC and structure formation for
\lapp1~Gyr. Forbes et al.\ (1997) suggest that some more localized
form of feedback from the first generation of stars and GCs could shut
off metal-poor GC formation for a few Gyr.  In the semi-analytic
simulations of Muratov \& Gnedin (2010), blue, metal-poor GCs are
formed in early mergers of low-mass protogalactic clumps, whereas both
blue and red GCs are formed in later mergers of more massive
galaxies. In any case, after the initial epoch of galaxy assembly and
GC formation, more GCs are expected to form each time the host galaxy
undergoes a gas-rich, dissipational major merger.  Because stellar
evolution enriches the interstellar and intergalactic medium over
time, any new GCs that are created in subsequent events are expected
to be more metal-rich than the earliest generation.  This type of
broad outline
can explain the metallicity gap between GC subpopulations of similar
age, as well as several other observations, such as:
more massive galaxies have larger relative numbers of blue, metal-poor
GCs (Rhode et al.\ 2005, 2007); the average metallicity of metal-poor
GCs in more luminous galaxies is slightly higher than that in less
luminous galaxies (Strader, Brodie, \& Forbes 2004); and blue,
metal-poor GCs have different spatial distributions than red,
metal-rich GCs
in some galaxies (e.g., Zinn 1985, Moore et al.\ 2006, Dirsch et
al.\ 2003, Bassino et al.\ 2006). On the other hand, this rough
outline ignores complicating effects such as dynamical destruction of
GCs over time, which may vary depending on the properties of the GC
itself and/or its parent galaxy (e.g., Vesperini 2000).
It also ignores the possibility that galaxy mergers may themselves
{\it destroy} a significant fraction of the GCs present in the
progenitor merging galaxies, not just create new ones (Kruijssen et
al.\ 2012).  Both of these effects may have substantially modified the
trends in GC numbers in different galaxies that we measure today (at
$z$$=$0) and may make it considerably more difficult to accurately
interpret our observations.

A scenario like this may also have implications for the observed
correlation between \ngc\ and \mbh. BT10 briefly speculate that the
apparent connection between 
\ngc\ and \mbh\ may arise because both SMBHs and GC populations grow
via recent major galaxy mergers.  Alternatively, SMBHs and GCs may be
linked because they originate at high-$z$ in gas-rich young galaxies.
We can consider these two possibilities in the context of the broad
picture I have outlined of 
the formation of
metal-poor and metal-rich GC subpopulations.  
If SMBHs are primordial objects and their properties are set during
the initial formation epoch of their host galaxies, one might expect a
strong correlation between the numbers of blue, metal-poor GCs in the
galaxy and the SMBH mass.  
If, on the other hand, SMBHs are fed primarily by recent major galaxy
mergers, then one might expect a stronger correlation between
\mbh\ and the numbers of red, metal-rich GCs in a galaxy, since the
metal-rich GC population is likely built up over time during major
merger events.

BT10 examined the \ngc$-$\mbh\ relation separately for blue and red
GCs in their sample of 13 galaxies but found that the correlations
were similar in quality; as they note, the fractions of blue and red
GCs in giant galaxies appear to be fairly constant. The blue fraction
$f_{\rm blue}$ for my sample of 20 galaxies is listed in column 11
of Table~\ref{table:master}.  The blue GC fractions for the galaxies
range from 0.52 to 0.72 and the mean $f_{\rm blue}$ is 0.61, with a
standard deviation of 0.07.  Figure~\ref{fig:blue red ngc smbh} shows
the log of the number of blue GCs and red GCs versus the log of the
SMBH mass plotted in the top panel and bottom panel,
respectively. (Galaxies with classical bulges and pseudobulges are
again shown with different symbols.)  The plots look similar and there
are only small differences in the fitting results from {\tt fitexy}.
Counting only the blue GCs, the best-fit relation for the 18 E/S0
galaxies in my sample
becomes:

\begin{equation}
\rm{{log\frac{M_{\rm SMBH}}{M_{\odot}} = (8.32\pm0.03) + (1.23\pm0.06)
    log\frac{N_{GC, blue}}{500}}}
\label{equation:ES0 bf blue}
\end{equation}

\noindent Including only the red GCs in my E/S0 galaxies yields:

\begin{equation}
\rm{{log\frac{M_{\rm SMBH}}{M_{\odot}} = (8.52\pm0.03) + (1.20\pm0.05)
    log\frac{N_{GC, red}}{500}}}
\label{equation:ES0 bf red}
\end{equation}

\begin{figure}
\includegraphics[width=3.6in]{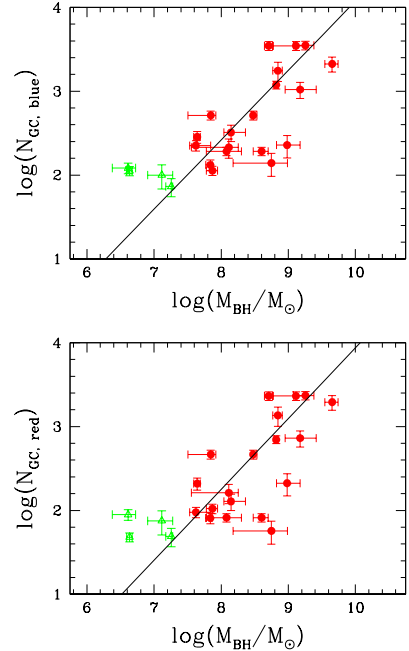}
\caption{\normalsize The log of \ngc\ versus the log of \mbh\ for the
  blue, metal-poor GCs (top) and red, metal-rich GCs (bottom) for the
  sample of 20 giant galaxies in this study.  Galaxies with classical
  bulges and pseudobulges are marked as in the other figures.
  NGC~1399, NGC~3379, and NGC~5128 each have two values of SMBH mass
  that are given half-weight in the fitting process. The solid lines
  are the best-fit lines derived using the 18 E/S0 galaxies in the
  sample (Equations~\ref{equation:ES0 bf blue} and \ref{equation:ES0 bf red}).}  
\protect\label{fig:blue red ngc
  smbh}
\end{figure}

\noindent These relations are shown as solid lines in the appropriate
plot (blue GCs in the top panel, red GCs in the bottom panel) in
Figure~\ref{fig:blue red ngc smbh}.  The blue-only GC sample yields a
slightly smaller reduced $\chi^2$ value than the red GC sample; the
former yields $\chi^2$/(N-4) $=$ 10.6 for the blue GCs
vs.\ $\chi^2$/(N-4) $=$ 11.0 for the red GCs. The blue GC sample also
needs slightly less additional scatter added to the data points than
the red GC sample in order to produce a reduced $\chi^2$ $\sim$ 1; the
blue GCs require $\epsilon_0$ $=$ 0.48 dex vs. $\epsilon_0$ $=$ 0.51
dex for the red GCs. Note that the blue GC fractions that appear in
Table~\ref{table:master} do not have accompanying uncertainties
listed.  Assigning an accurate error to the blue fraction can be
difficult, especially when data are being compiled from different
studies that use a variety of methods for determining blue and red GC
fractions (e.g., fitting with mixture modeling code, or simply
dividing a completeness-corrected sample at a certain fiducial color)
and in many cases no error is given on the estimated blue or red fraction.
To investigate how possible uncertainties on the blue and red fraction
would contribute to the relation between number of blue or red GCs and
SMBH mass, I incorporated a 10\% error on the blue and red fractions
into the total errors on $N_{\rm GC, blue}$ and $N_{\rm GC, red}$ and
ran {\tt fitexy} again for the 18 E/S0 galaxies.  Including only the
blue GCs, and incorporating a 10\% error on the blue fractions given
in Table~\ref{table:master}, the best-fit relation for the 18 E/S0
galaxies is:

\begin{equation}
\rm{{log\frac{M_{\rm SMBH}}{M_{\odot}} = (8.32\pm0.03) + (1.29\pm0.07)
    log\frac{N_{GC, blue}}{500}}}
\label{equation:ES0 bf blue with errors}
\end{equation}

\noindent Including only the red GCs and including a 10\% error on the
red fraction yields for the E/S0 galaxies:

\begin{equation}
\rm{{log\frac{M_{\rm SMBH}}{M_{\odot}} = (8.52\pm0.04) + (1.23\pm0.06)
    log\frac{N_{GC, red}}{500}}}
\label{equation:ES0 bf red with errors}
\end{equation}

\noindent With these additional error estimates included, the reduced
$\chi^2$ for the fit to the blue GC sample is $\chi^2$/(N-4) $=$ 9.0
and the fit to the red GC sample yields $\chi^2$/(N-4) $=$ 9.5.

If the blue GC population were a
better tracer of SMBH mass than the red GC population {\it and} if the
overall picture of hierarchical galaxy and GC formation actually holds
true, it might
suggest that the masses of the SMBHs are set during the
initial formation epoch of the host galaxies and are less dependent on
subsequent mergers that occur.  However, given that reduced $\chi^2$
values are themselves subject to uncertainty (on the order of a few
tenths in this case; e.g., Andrae et al.\ 2010), 
the very small difference in reduced $\chi^2$ between the blue and red
GC populations is not significant.

\subsection{Trends with Host Galaxy Morphology}

\subsubsection{Splitting the Sample by Hubble Type}
\protect\label{section:hubble type}

\begin{figure}
\includegraphics[width=3.6in]{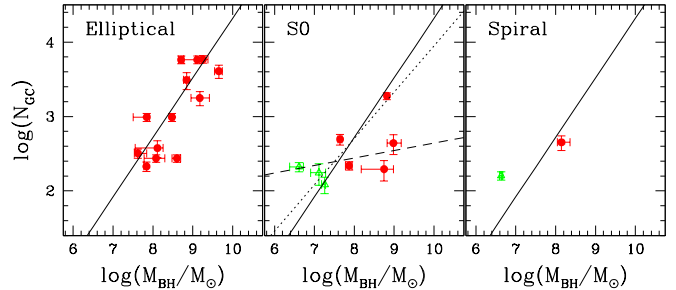}
\caption{\normalsize The same data as in Figure~\ref{fig:ngc smbh},
  except with the 20 galaxies in the final sample separated by
  morphological type. Red filled circles again represent galaxies with
  classical bulges and green open triangles show galaxies with
  pseudobulges.  The solid line is the best-fit \ngc$-$\mbh\ relation
  for E/S0 galaxies (Equation~\ref{equation:ES0 bf}). NGC~4594 is
  included in the S0 plot, although this galaxy is sometimes
  classified as an Sa. The dotted line in the middle panel is the
  best-fitting relation for the S0s with NGC~4594 included, and the
  dashed line is the best-fitting relation with NGC~4594 excluded.}
\protect\label{fig:morph ngc smbh}
\end{figure}

HH11 compiled a larger sample of galaxies than BT10 with a wider range
of morphological types and hence were able to investigate how the
correlation of \ngc\ with \mbh\ varies with galaxy morphology. The
HH11 sample included 21 elliptical galaxies, eight S0 galaxies, and
four spiral galaxies.  My more restricted sample of 20 galaxies with
reliable GC system properties includes 10 ellipticals, eight
lenticular galaxies, and two spiral galaxies, so I can also explore
this issue, although to a more limited extent than HH11.
Figure~\ref{fig:morph ngc smbh} shows log~\ngc\ versus log~\mbh\ for
the 20 galaxies
separated by morphological type, with ellipticals, S0 galaxies, and
spiral galaxies plotted from left to right.
The best-fit \ngc$-$\mbh\ relation for E/S0s
(Equation~\ref{equation:ES0 bf}) is shown as a fiducial; it appears in
all three plots as a solid line.  The three galaxies with two
\mbh\ values, NGC~1399, NGC~3379, and NGC~5128 are (as before) plotted
twice; all three are Es and thus are included in the leftmost plot.
The morphological type of NGC~4594 (M104, a.k.a. the Sombrero galaxy)
is listed in Table~\ref{table:master} as ``S0/Sa'' and I have included
this galaxy in the S0 plot; it is the point with the largest
\ngc\ value, lying very close to the E/S0 best-fit line.  NGC~4594 is
often classified by appearance as an Sa spiral galaxy, probably at
least in part because its nearly edge-on orientation makes its dust
lane and disk appear more prominent.
Its quantitative properties, however, are more like that of an S0
galaxy: its bulge-to-disk ratio is $\sim$6, bulge fraction ($B/T$) is
0.86 (Kent 1988), and its $B-V$ color is 0.84 (RC3; de~Vaucouleurs et
al.\ 1991). In any case, it is unclear whether NGC~4594 belongs in the
S0 plot (middle) or the spiral galaxy plot (right) in
Figure~\ref{fig:morph ngc smbh}.

HH11 found that the elliptical galaxies in their sample exhibited a
strong correlation between \ngc\ and \mbh\ and that the spiral
galaxies, with the exception of the Milky Way, closely followed the
same relation. The S0 galaxies in their sample, on the other hand,
showed ``no clear trend'' in a log(\ngc) versus log (\mbh) plot.  The
results for my sample of 20 galaxies are similar to those of HH11.
The 10 elliptical galaxies in the left panel of Figure~\ref{fig:morph
  ngc smbh} closely follow the fiducial E/S0 line.  When I fit the
\ngc$-$\mbh\ relation for just these 10 elliptical galaxies, the
result is

\begin{equation}
\rm{{log\frac{M_{\rm SMBH}}{M_{\odot}} = (8.17\pm0.04) + (0.95\pm0.07)
    log\frac{N_{GC}}{500}}}
\end{equation}

\noindent with a reduced $\chi^2$ values of $\chi^2$/(N$-$4) $=$ 9.4
and $\chi^2$/(N$-$2) $=$ 7.7 and $\epsilon_0$ $=$ 0.37.  (For
comparison, the slope and intercept of the elliptical-only relation
from HH11 are 0.98$\pm$0.10 and 8.30$\pm$0.29, respectively; see
Equation~\ref{equation:HH11 E bf}.)  The slope and intercept of this
line are the same within the errors as those for my E/S0 sample
(Equation~\ref{equation:ES0 bf}).  The spiral galaxy points in the
rightmost panel in the figure both fall close to the E/S0 line. In the
middle panel of S0 galaxies, NGC~4594 (which may be an Sa or an S0
galaxy) also falls close to the E/S0 best-fit line.  The other S0
galaxies have a relatively small range of \ngc\ values
($\sim$120$-$500) but a large range in SMBH mass.  Fitting a line to
the S0 points with NGC~4594 included changes the slope of the best-fit
\ngc$-$\mbh\ relation to 1.6$\pm$0.1.  Excluding NGC~4594 and fitting
the points for the remaining S0 galaxies yields only a very shallow
trend of slightly increasing log(\ngc) with increasing log(\mbh) with
a slope of 10.2$\pm$5.5; this is not consistent with the best-fitting
relations for the E/S0 sample or the elliptical-only sample.

\subsubsection{Classical Bulges vs.\ Pseudobulges}
\label{section:pseudo classical}

Before drawing any conclusions about trends in morphology, it is worth
exploring one more morphology-related issue and asking whether the
behavior of galaxies with classical bulges differs from that of
galaxies with pseudobulges in the \ngc$-$\mbh\ plane.
%
Classical bulges are thought to form via relatively rapid (i.e.,
evolution on dynamical time scales), violent processes like
dissipational collapse and mergers.  In contrast, pseudobulges appear
to be built up gradually over a few billion years by ``secular'', internal
processes in galaxy disks, like gas transport aided by bar
instabilities and spiral structure \citep{kk04}. By this definition,
elliptical galaxies are in the ``classical'' bulge category and S0 and
spiral galaxies can fall in either category, although \citet{kk04}
note that the type of secular evolution that produces pseudobulges is
most likely to be important in intermediate- to late-type galaxies
(Sbc galaxies and later).
Kormendy, Bender, \& Cornell (2011) recently analyzed a sample of
$\sim$50 spiral, S0, and elliptical galaxies and found that the disk
galaxies with pseudobulges do not follow the same $M$-$\sigma$
relation as ellipticals and classical-bulge galaxies. Galaxies in
their sample with pseudobulges at least showed a much larger scatter
in the $M$-$\sigma$ plane and arguably displayed no correlation at
all.

To examine whether the \ngc$-$\mbh\ correlation changes for galaxies
with pseudobulges vs.\ those with classical bulges, I searched the
literature to find bulge classifications for the 20 galaxies in my
sample. 
The elliptical galaxies 
all fall in the ``classical'' category.  
I found classifications for the S0 and spiral galaxies 
primarily in the lists included in Kormendy \& Kennicutt (2004),
Fisher \& Drory (2008, 2010), and Kormendy et al.\ (2011). Based on my
literature search, four of the 20 galaxies in my sample {\it may} have
pseudobulges: the Milky Way (Shen et al.\ 2010 and references
therein), NGC~3384 (Kormendy et al.\ 2011, Pinkney et al.\ 2003),
%
NGC~7332 (Pinkney et al.\ 2003, Falcon-Barroso et al.\ 2004), and
NGC~7457 (Kormendy 1993, Kormendy \& Kennicutt 2004).  
Classifying a galaxy as possessing a pseudobulge is
notoriously difficult (Kormendy \& Kennicutt 2004), so 
none of these four galaxies has been {\it unequivocally} shown to have
a pseudobulge rather than a classical bulge.
The classifications of NGC~7332 and NGC~7457 are especially uncertain.
The evidence that NGC~7332 has a pseudobulge is merely circumstantial:
observations with an integral-field unit show that this galaxy has a
bar and a cold, counter-rotating stellar disk, as well as stellar
kinematics in its inner regions that are typical of a ``boxy bulge''
(Falcon-Barroso et al.\ 2004).  Together these imply a pseudobulge
rather than a classical bulge. In the case of NGC~7457, Kormendy
(1993) found that its central velocity dispersion is very small for
its bulge luminosity (i.e., its bulge lies well below the
Faber-Jackson relation; Faber \& Jackson 1976); this is one of the
properties Kormendy \& Kennicutt (2004) identify with pseudobulges.
Pinkney et al.\ (2003) confirmed that NGC~7457 has central stellar
kinematics characteristic of a pseudobulge, but Kormendy et
al.\ (2011) place this galaxy in the ``classical bulge'' category in
their paper.

The four galaxies that may have pseudobulges are designated in
Figures~\ref{fig:ngc smbh}$-$\ref{fig:morph ngc smbh} with green open
triangles, while the classical-bulge and elliptical galaxies are
plotted with red filled circles.  Figure~\ref{fig:ngc smbh} shows the
full sample of 20 galaxies in the \ngc$-$\mbh\ plane; the four
pseudobulge galaxies are in the lower left of the diagram. In
Figure~\ref{fig:morph ngc smbh} the galaxies are split up by
morphology, so three of the pseudobulge galaxies (NGC~3384, NGC~7332,
and NGC~7457) appear in the middle panel and the fourth (the Milky
Way) appears on the spiral galaxy panel on the right. Three of the
four pseudobulge galaxies are systematically above and to the left of
the best-fit E/S0 relation in the \ngc$-$\mbh\ plane, i.e., their GC
populations are too large for their SMBH masses (or their SMBH masses
are too small for their GC numbers).  The other pseudobulge galaxy
(NGC~3384) lies on the relation. On the other hand, there are so few
pseudobulge points and there is so much scatter around the
best-fitting \ngc$-$\mbh\ relation
that it is not possible to determine from this sample whether the
pseudobulge galaxies are truly outliers or are simply, by coincidence,
scattering in the same direction.

Lastly, I should note that the galaxies with pseudobulges seem to fall
along the same $M$-$\sigma$ relation (Figure~\ref{fig:m sigma}) as the
rest of the galaxies in the full sample.  Two of the four galaxies
with pseudobulges (NGC~3384 and NGC~7332) intersect the best-fitting
line for the full sample (the solid line in Figure~\ref{fig:m sigma}),
and a third galaxy (the Milky Way) lies very close to the line.  Only
NGC~7457 scatters far away from the best-fit relation (and its
classification as a pseudobulge galaxy was uncertain).
As mentioned earlier, Kormendy et al.\ (2011) concluded that
pseudobulge galaxies do not follow the $M$-$\sigma$ relation for
ellipticals and classical-bulge galaxies. The galaxies that seemed to
deviate from their best-fitting $M$-$\sigma$ relation in Kormendy et
al.\ were located in a specific region of the $M$-$\sigma$ plane:
below SMBH masses log(M/$M_\odot$) \lapp8 and velocity dispersions
log($\sigma$) \lapp2.2.  Kormendy et al.\ (2011) had $\sim$15 galaxies
with pseudobulges in that area of the plane. My sample has very few
objects in that region of the $M$-$\sigma$ plane so it is difficult to
assess whether the data presented here contradict their conclusions or
not.

\subsubsection{What Might the Trends in Morphology Mean?}
\protect\label{section:morphology interpretation}

%
In part because of the small numbers of S0, spiral, and pseudobulge
galaxies in the sample, it is not clear what the pseudobulge
vs.\ classical bulge values in the \ngc$-$\mbh\ plane and the data in
Figure~\ref{fig:morph ngc smbh} are indicating.
It is possible
that there is: (1) a marked trend of increasing \ngc\ with increasing
\ \mbh\ mass for elliptical galaxies and perhaps spiral galaxies, but
a much weaker trend for S0s; or (2) a clear trend in increasing
\ngc\ with \mbh\ for classical-bulge galaxies but not for pseudobulge
galaxies; or (3) an overall trend in \ngc\ vs.\ \mbh\ for giant
galaxies, but with plenty of scatter, especially at lower \ngc\ and
lower SMBH masses.

The situation implied by the first possibility does not have an
obvious physical explanation; it is not clear what mechanism would
cause 
S0 galaxies to have a limited range of \ngc\ values but a wide range
of SMBH masses (log(\mbh/M$_\odot$) $\sim$ 6$-$9) while ellipticals
have \ngc\ and \mbh\ values that are closely linked.  Whatever direct
or indirect correlation might be present between GC populations and
SMBH masses in elliptical galaxies may, for some unknown reason, break
down for most S0 galaxies.

The second possibility, on the other hand, could make sense in terms
of how we think both SMBHs and bulges grow. Kormendy et al.\ (2011)
succinctly describe two different ``modes'' for the feeding and growth
of SMBHs, that echo the two different growth modes for building galaxy
bulges.  One mode is violent, rapid mergers that drive gas infall and
efficiently build up both classical bulges and the seed SMBHs within
them.  The other mode is the more gradual, local, stochastic growth
(e.g., build-up of a central concentration in a disk galaxy via disk
instabilities and gas-flow along bars) that builds pseudobulges, and
that may also result in slower and more modest growth of SMBHs.  If GC
populations, classical bulges, and SMBHs can all be grown and
significantly increased via mergers, one might imagine that galaxies
with classical bulges are then also likely to have \ngc\ values and
\mbh\ values that increase together.
%
Galaxies with more quiescent histories, that perhaps made GCs in their
early assembly stages at high-$z$ but did not do much more to
supplement their GC populations over time, perhaps would end up today
as disk galaxies with a pseudobulge, a modest-sized SMBH, and a weaker
correlation between their \mbh\ and \ngc\ values.
%
In other words, whatever correlation between \ngc\ and \mbh\ was set
up in the earliest stages of a galaxy's assembly may be strengthened
in galaxies that continue to grow via major mergers, but weakened in
galaxies with pseudobulges that evolve more quiescently.

Finally, the third 
scenario mentioned above is that the \ngc$-$\mbh\ correlation is
generally present in giant galaxies but is weaker and has more scatter
at lower \ngc\ values and lower SMBH masses, and by implication,
lower-mass galaxies, since both \ngc\ and \mbh\ correlate with galaxy
mass (see next section).
If this is the case, the explanation could be similar to the scenario
for possibility (2).  Perhaps a common mechanism like major galaxy
mergers drives the formation of very massive SMBHs and very populous
GC systems in massive galaxies
and this mechanism becomes less important in the evolution of
lower-mass galaxies, their GC populations, and their SMBHs.  The net
effect would be that the GC numbers and SMBHs ``decouple'' (because
the phenomenon that linked them in the first place, major gas-rich
mergers, is no longer of primary importance in their evolution) and
the correlation gets weaker with decreasing galaxy mass.

\subsection{Correlations of \ngc\ with Other Galaxy Properties}
\protect\label{section:ngc correlations}

\subsubsection{Number of GCs Versus Galaxy Stellar Luminosity and Mass}
\protect\label{section:ngc lum mass}

The number of GCs in some giant galaxies has been observed to follow a
rough correlation with galaxy luminosity, with \ngc\ $\propto$
$L^{\alpha}$, where $\alpha$ is between 1 and 2 (Ashman \& Zepf 1998
and references therein).  A sensible next step is to use the current
galaxy sample to examine this and other correlations that \ngc\ has
with galaxy properties,
as well as to compare the strength of this correlation to the
\ngc$-$\mbh\ relation. To do this, I have supplemented the galaxies in
the N$=$20 sample that I used in earlier sections of the paper with
ten more galaxies with measured \ngc\ from
my wide-field GC system survey.
These ten additional galaxies have well-determined GC system
properties (i.e., the criteria in 
Section~\ref{section:criteria} are met), but the galaxies were not
included in the previous analysis because they do not have measured
SMBH masses.  The properties of these additional ten galaxies are
listed in Table~\ref{table:wfgcss}. The columns in the table are:
galaxy name and morphological type; predicted SMBH mass calculated
from the best-fitting \ngc$-$\mbh\ relation
(Equation~\ref{equation:all bf}; see Section~\ref{section:wfgcss m
  sigma} for details); central velocity dispersion; $M_V$ for the
galaxy; distance to the galaxy in Mpc; number of GCs; GC specific
frequencies $S_N$ and $T$; fraction of blue GCs in the GC system; and
the reference for the GC system properties.

\begin{figure}

\includegraphics[width=3.2in]{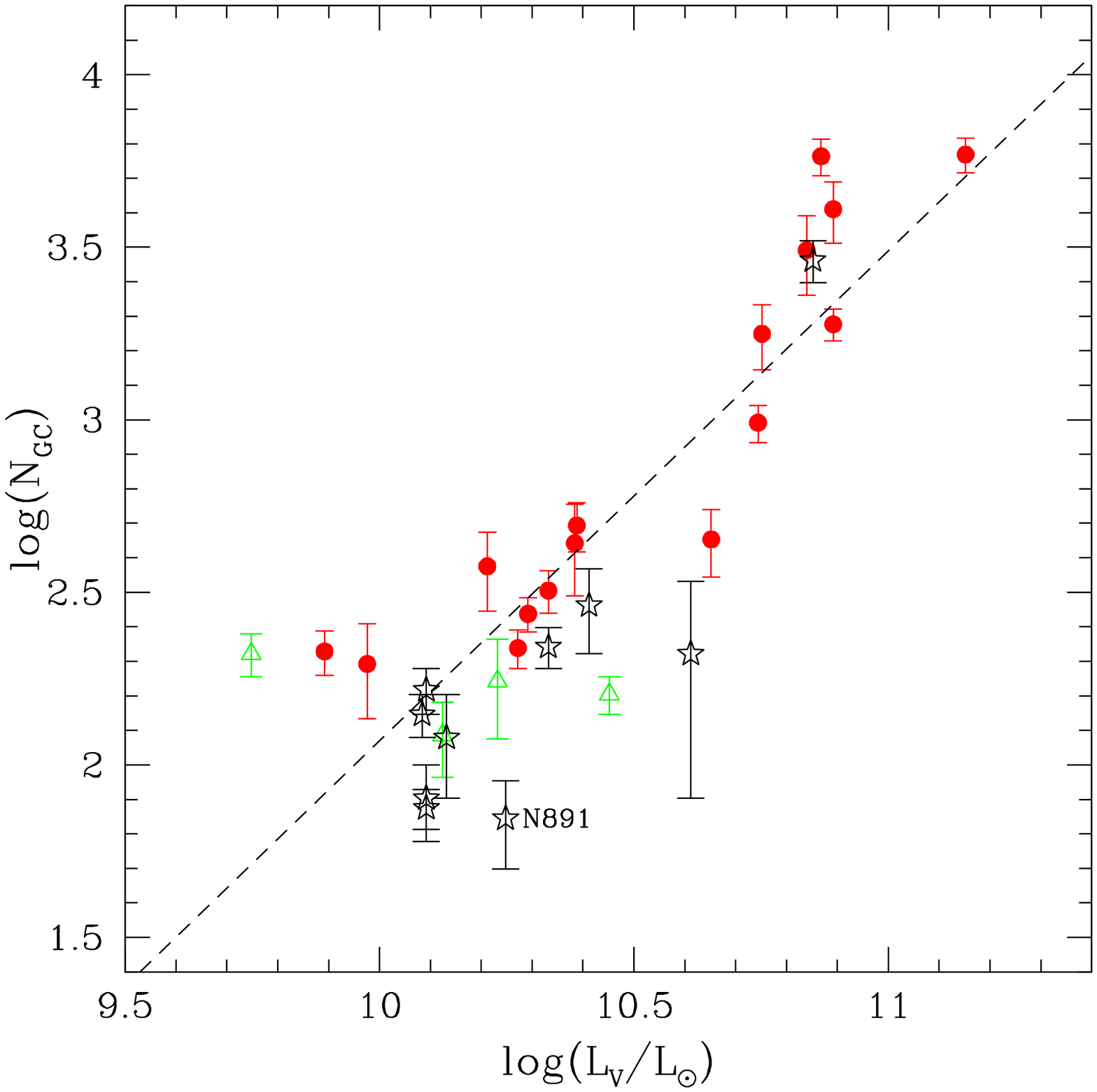}\hskip 0.05in \includegraphics[width=3.2in]{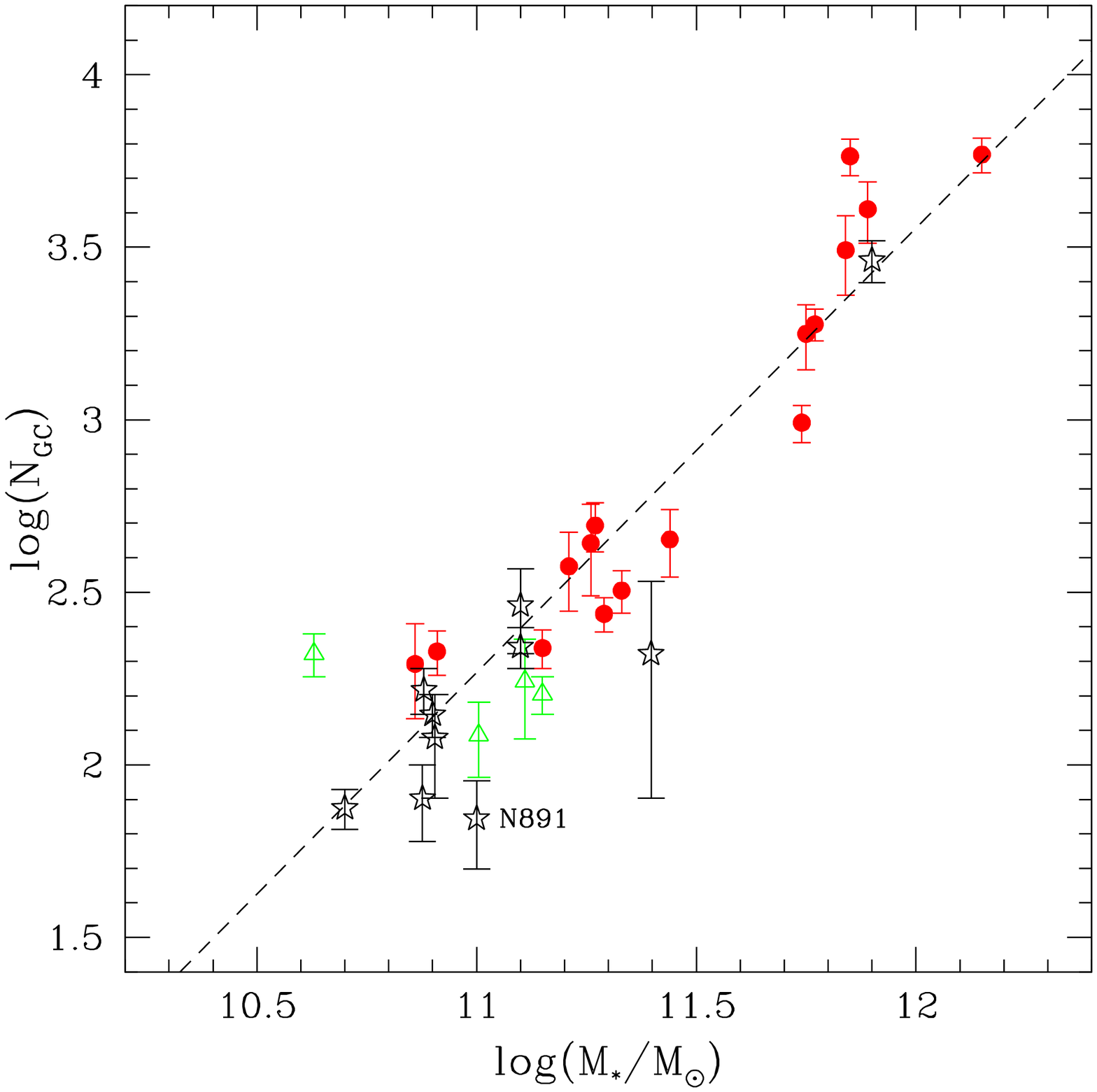}
\caption{\normalsize The log of \ngc\ versus the log of the 
  $V$-band luminosity 
and galaxy stellar mass 
for the 20 galaxies in Table~\ref{table:master} 
  and the ten galaxies from my wide-field survey in
  Table~\ref{table:wfgcss}.  Galaxies with classical bulges and
  pseudobulges in the Table~\ref{table:master} sample are marked with
  red filled circles and green open triangles, respectively, as in the
  other figures.  Galaxies from the wide-field survey are designated
  with open stars.  The spiral galaxy NGC~891 (labeled on the plots)
 is not shown in green but may have a pseudobulge.  The best-fit lines
 are plotted on the figures and given in Equations~\ref{equation:bf
   ngc lum} and \ref{equation:bf ngc galmass}. The relation between
 \ngc\ and total galaxy luminosity yields $\chi^2/(N-2)$ $=$ 15.8 and
 the relation between \ngc\ and total galaxy stellar mass yields
 $\chi^2/(N-2)$ $=$ 8.9.}  
\protect\label{fig:ngc lum galmass}
\end{figure}

The left-hand side of Figure~\ref{fig:ngc lum galmass} shows the log
of \ngc\ versus the log of the $V$-band luminosity in solar units,
$L_V$/$L_{\odot}$, for the 30 galaxies listed in
Tables~\ref{table:master} and \ref{table:wfgcss}.  The right-hand side
of the same figure shows log of \ngc\ plotted against the log of the
galaxy stellar mass in solar masses ($M_*$/$M_{\odot}$) for the same
30 galaxies.  To calculate stellar masses for the galaxies, I combined
their $M_V$ values with the Zepf \& Ashman (1993) mass-to-light ratios
discussed in Section~\ref{section:criteria}.  In both plots, galaxies
from Table~\ref{table:master} with classical bulges are designated
with red filled circles and galaxies with pseudobulges with green open
triangles, while the 10 additional galaxies from the wide-field survey
are plotted with open stars.  I searched the literature for bulge
classifications for these additional 10 galaxies and found that one of
the objects, the Sb spiral galaxy NGC~891, {\it may} have a
pseudobulge; de~Jong et al.\ (2008) characterize its bulge as a boxy
pseudobulge based on surface photometry. The point representing
NGC~891 is labeled in the plots.
The point with the largest \ngc\ error bars in both plots is NGC~7331; this
galaxy has a relatively low inclination ($i$$\sim$75~degrees) compared
to the other spiral galaxies in the survey, which makes it difficult
to detect GCs against the background of the galaxy's spiral disk.  The
end result was a \ngc\ value that was relatively uncertain (Rhode et
al.\ 2007).  Lastly, note that there are only 30 points in the figure
because the three galaxies that were plotted twice in the figures that
included SMBH mass are only plotted once here (they each have only one
value of \ngc\ and galaxy luminosity or stellar mass).

Fitting the 30 data points shown in the panels of Figure~\ref{fig:ngc
  lum galmass}
yields the following best-fit linear relations:
\begin{equation}
\rm{{log\frac{L_{\rm V}}{L_{\odot}} = (10.44\pm0.01) + (0.70\pm0.02)
    log\frac{N_{GC}}{500}}}
\protect\label{equation:bf ngc lum}
\end{equation}

\noindent and

\begin{equation}
\rm{{log\frac{M_{*}}{M_{\odot}} = (11.33\pm0.01) + (0.78\pm0.02)
    log\frac{N_{GC}}{500}}}
\protect\label{equation:bf ngc galmass}
\end{equation}

\noindent Written another way, these relationships mean
\ngc\ $\propto$ $(L_V$/$L_{\odot})^{1.42}$ and \ngc\ $\propto$
$(M_*$/$M_{\odot})^{1.28}$, respectively.  The reduced $\chi^2$ value
for the fit between \ngc\ and total galaxy stellar mass is
$\chi^2$/$(N-2)$ $=$ 8.9, compared to 15.8 for the \ngc$-$luminosity
fit.  This makes sense given the appearance of the data: the log of
the number of GCs shows a tight correlation with the log of the host
galaxy stellar mass, whereas the log(\ngc) vs.\ log($L_V$) data show
much more scatter, especially for the lower-luminosity galaxies with
log($L_V$) $<$ 10.75.  The intrinsic scatter is substantially smaller
in this relation compared to any of the \ngc$-$\mbh\ relations: adding
$\epsilon_0$ $=$ 0.14 dex in quadrature to the errors brings the
reduced $\chi^2$ value to $\sim$1.  (For completeness, I also checked
the fitting result for the 20-galaxy sample in
Table~\ref{table:master}, without the additional ten galaxies from the
survey.  For the N$=$20 sample, the intrinsic scatter remains small,
at $\epsilon_0$ $=$ 0.15~dex.) The correlation between \ngc\ and
galaxy stellar mass in Figure~\ref{fig:ngc lum galmass} is the
strongest of any of the correlations analyzed in this paper. It is
important to note that the reduced $\chi^2$ and $\epsilon_0$ values
would be even smaller if errors on the galaxy luminosity or mass were
included in the fits.  Specifically, arbitrarily setting the galaxy
mass error to 10\% of the stellar mass yields a best-fitting line with
the same slope and intercept (within the uncertainties on those
coefficients) as the fit without errors on the mass, but with a
$\chi^2$/$(N-2)$ $=$ 5.1.  Setting the relative error on the galaxy
stellar mass to 25\% yields the same coefficients and $\chi^2$/$(N-2)$
$=$ 1.5.  Even without error bars in the $x$-direction, the quality of
the correlation between \ngc\ and galaxy mass is much better than the
\ngc$-$\mbh\ and $M$-$\sigma$ correlations explored in the earlier
sections of the paper.

When I include only the blue, metal-poor or red, metal-rich GCs in
\ngc\ and fit the data points versus total galaxy stellar mass, I find
that the slope of the line remains the same within the errors, but
that the quality of the fit is better for the blue GCs than for the
red ones.  The former yields $\chi^2$/$(N-2)$ $=$ 8.7 and $\epsilon_0$
$=$ 0.14~dex, whereas the latter yields $\chi^2$/$(N-2)$ $=$ 11.1 and
$\epsilon_0$ $=$ 0.15~dex.  The slightly improved fit yielded by the
blue GCs may suggest that it is chiefly the blue, metal-poor GCs that
are driving the correlation between \ngc\ and galaxy stellar mass.

One object in Figure~\ref{fig:ngc lum galmass}, the S0 galaxy
NGC~7457, has a particularly large value of \ngc\ for its stellar
mass, and lies significantly off the best-fit relation. It is the
lowest-mass galaxy in the sample, with log($M_*$/$M_{\odot}$) $=$ 10.6
but it has an estimated population of 210$\pm$30 GCs (Hargis et
al.\ 2011). This galaxy is in a low-density environment ($\rho$ $=$
0.13 Mpc$^{-3}$; Tully 1988) and its GC system was studied as part of
our wide-field survey. There is not an obvious reason for its high
\ngc\ value and high specific frequency ($S_N$ $=$
3.1$\pm$0.7). Excluding NGC~7457 from the sample and fitting log(\ngc)
vs.\ log($M_*$/$M_{\odot}$) yields the same coefficients as in
Equation~\ref{equation:bf ngc galmass} within the errors and improves
the fit, so that $\chi^2$/$(N-2)$ becomes 6.0 and $\epsilon_0$ becomes
0.10~dex.

Lastly, I should note that in addition to calculating total galaxy
stellar masses from the $V$-band luminosity, I also computed stellar
masses from $K$-band luminosity of each galaxy and used those to
examine the relationship between \ngc\ and galaxy stellar mass.  I
combined the total $K$-band apparent magnitude listed in the NASA
Extragalactic Database (NED) --- which in turn come from either from
the 2MASS Extended Source Catalog (Jarrett et al.\ 2000) or the 2MASS
Large Galaxy Atlas (Jarrett et al.\ 2003) --- with the adopted
distance modulus to derive total $K$-band luminosity for each
galaxy. I then applied the appropriate $K$-band mass-to-light ratio
from Spitler et al.\ (2008) for each galaxy's morphological type to
calculate galaxy stellar mass.
I fitted log(\ngc) vs.\ log($M_*$/$M_{\odot}$) for these
$K$-band-derived masses and found that the best-fitting coefficients
were similar to those in Equation~\ref{equation:bf ngc galmass} but
the scatter in the plot was larger and the reduced $\chi^2$ values
increased by a factor of $\sim$1.5.  So at least for this specific
data set, the $V$-band galaxy magnitudes and stellar masses produced
better-quality results in terms of the correlation between number of
GCs and galaxy stellar mass.

\subsubsection{Number of GCs Versus Bulge Luminosity and Mass}
\protect\label{section:ngc bulge lum mass}

The analysis in the previous section examined correlations of
\ngc\ with the total stellar mass and luminosity of the host galaxy,
but there are reasons why looking at \ngc\ versus the mass and
luminosity of just the spheroidal component might make more sense.  As
mentioned in Section~\ref{section:introduction}, the analysis of
Snyder et al.\ (2011) indicated that the correlation between \ngc\ and
\mbh\ arises because both of those quantities correlate with the
binding energy or potential well depth of the bulge component of the
host galaxy. It has been argued that at least some GCs are actually
associated with galaxy bulges rather than with thick disks or halos.
Specifically, the properties of the more metal-rich GC subpopulation
of the Milky Way may indicate it is a bulge population
(e.g., van den Bergh 1993).  In this type of picture, the red,
metal-rich GCs would then be an analogous population in other spiral
and elliptical galaxies (e.g., Forbes, Brodie \& Larsen 2001).  In
previous extragalactic GC system studies, the specific frequency $S_N$
is sometimes normalized using bulge luminosity rather than total
luminosity for spiral galaxies.  (e.g., Harris 1981, Cote et al. 2000,
Forbes et al.\ 2001).  For these reasons, I have also investigated the
relationship between \ngc\ and galaxy luminosity and mass using the
{\it bulge} luminosity and mass of the spiral galaxies rather than the
total stellar luminosity and mass.

To calculate bulge luminosities and masses for the sample galaxies, I
found published estimates of the mean bulge fraction ($B/T$, the
fraction of the total luminosity that comes from the bulge) for
galaxies of various Hubble types.  I used the data given in Binney \&
Merrifield (1998), which is based on the work of Kent (1985), and
combined it with the total $V$-band magnitudes and galaxy stellar
masses in Tables~\ref{table:master} and \ref{table:wfgcss} to
calculate $M_V$ and log($M$/$M_{\odot}$) for each galaxy's bulge.  I
assumed bulge fractions of 1.0 for the elliptical galaxies in the
sample, 0.65 for the lenticular galaxies, 0.375 for the Sab galaxies,
0.25 for the Sb galaxies, 0.2 for the Sbc galaxies, and 0.15 for the
Sc galaxy. For NGC~4594, I used the $B/T$ measurement from Kent (1988)
that was mentioned in Section~\ref{section:hubble type}. I also tried
using bulge fractions derived from data given in Simien \&
de~Vaucouleurs (1986), which vary from the Binney \& Merrifield values
by a few tenths, and finally I tried applying the bulge-to-disk ratios
for spiral galaxies given in Graham \& Worley (2008).  For both the
Simien \& de~Vaucouleurs (1986) bulge-to-disk ratios and the Graham \&
Worley (2008) ratios, the resultant \ngc\ vs. bulge luminosity and
bulge mass fits yielded similar slopes and intercepts but produced
slightly more scatter and higher $\chi^2$ values.  I have adopted the
Binney \& Merrifield bulge fractions as the final values.

The left panel in Figure~\ref{fig:ngc lum galmass bulge} shows the log
of \ngc\ versus the log of the $V$-band luminosity of the host galaxy
bulge for the 30 galaxies in Tables~\ref{table:master} and
\ref{table:wfgcss}.  The symbols are the same as in
Figure~\ref{fig:ngc lum galmass}.  The positions of the data points
for the elliptical galaxies have not changed compared to
Figure~\ref{fig:ngc lum galmass}, but the spiral and S0 galaxies (in
the lower-luminosity region of the diagram) show smaller scatter and a
more linear behavior.  For example, NGC~7457 was the galaxy with the
lowest luminosity in Figure~\ref{fig:ngc lum galmass} and was an
outlier, whereas in Figure~\ref{fig:ngc lum galmass bulge} it lies
within the cluster of other data points at the low-luminosity end of
the sample.  The right-hand panel in Figure~\ref{fig:ngc lum galmass
  bulge} shows log of \ngc\ plotted against the mass of the galaxy
bulge component.  Some of the outliers in Figure~\ref{fig:ngc lum
  galmass} --- specifically, NGC~7457 and NGC~7331 --- also have
become less discrepant in this figure compared to the plot of
log(\ngc) versus total galaxy mass.

\begin{figure}
\includegraphics[width=3.2in]{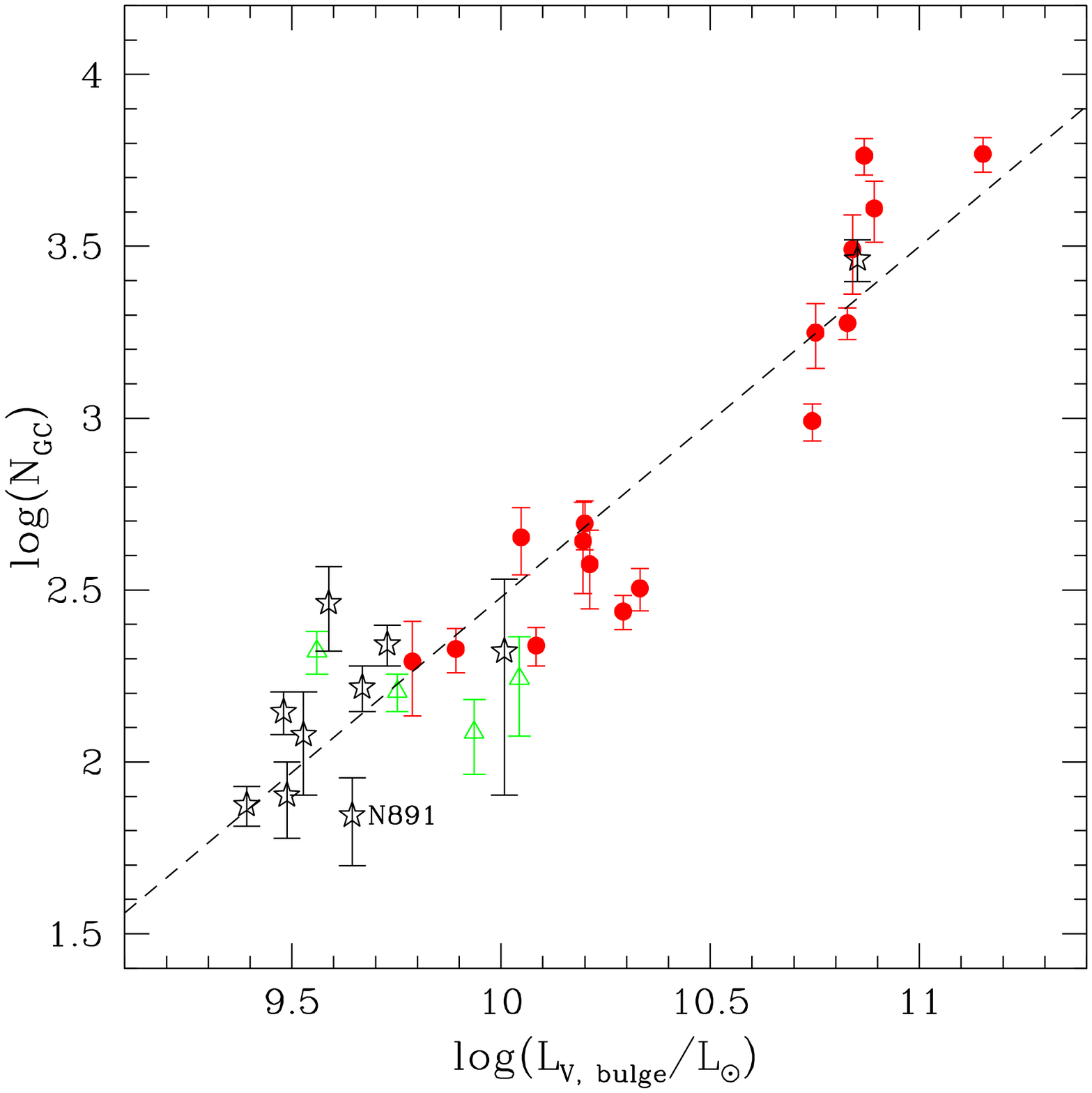}\hskip 0.05in \includegraphics[width=3.2in]{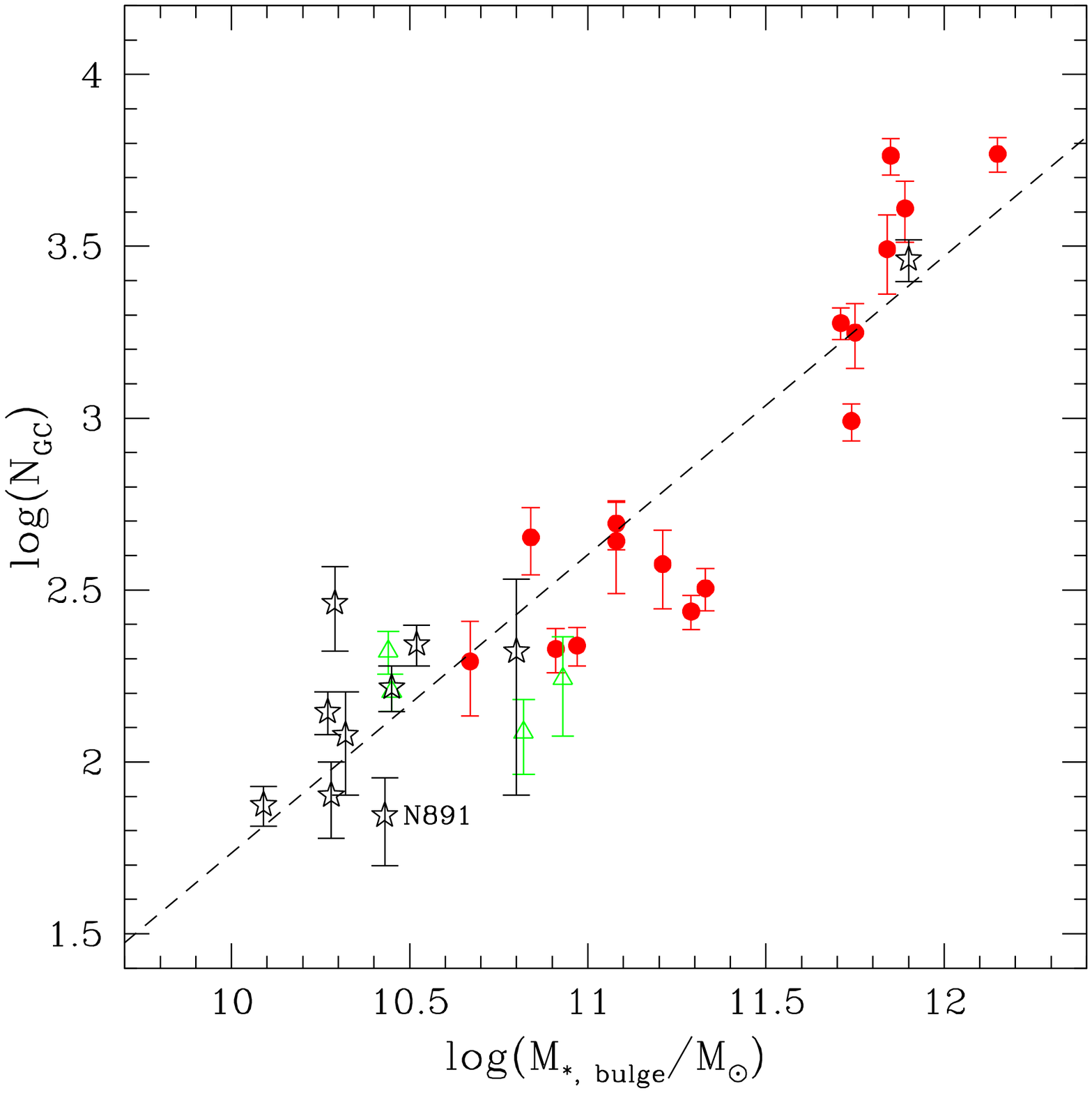}
\caption{\normalsize The log of \ngc\ versus the log of the $V$-band
  luminosity 
or stellar mass 
of the bulge for spiral galaxies or the entire galaxy for elliptical
galaxies.  The data points represent the 30 galaxies in
Tables~\ref{table:master} and \ref{table:wfgcss}. The symbols are the
same as in Figure~\ref{fig:ngc lum galmass}. The spiral galaxy NGC~891
is not shown in green but may have a pseudobulge. The best-fitting
relations are shown as a dashed lines and given in
Equations~\ref{equation:bf ngc lum bulge} and \ref{equation:bf ngc
  galmass bulge}. The relation between \ngc\ and bulge luminosity
gives $\chi^2/(N-2)$ $=$ 9.2 and the relation between \ngc\ and bulge
mass gives $\chi^2/(N-2)$ $=$ 11.5.}  \protect\label{fig:ngc lum
  galmass bulge}
\end{figure}

The quantitative results from the fitting process are mixed. The log
of \ngc\ correlates much more closely with the log of the bulge
luminosity than it does with total galaxy luminosity.  However, the
correlation of log(\ngc) with bulge mass is much weaker than that with
total galaxy stellar mass.  The best-fitting lines to the 30 data
points in Figure~\ref{fig:ngc lum galmass bulge} are

\begin{equation}
\rm{{log\frac{L_{\rm V, bulge}}{L_{\odot}} = (10.22\pm0.01) + (0.98\pm0.02)
    log\frac{N_{GC}}{500}}}
\protect\label{equation:bf ngc lum bulge}
\end{equation}

\noindent and

\begin{equation}
\rm{{log\frac{M_{*, bulge}}{M_{\odot}} = (11.11\pm0.01) + (1.15\pm0.03)
    log\frac{N_{GC}}{500}}}
\protect\label{equation:bf ngc galmass bulge}
\end{equation}

\noindent which translate to the proportionalities \ngc\ $\propto$
$(L_{\rm V, bulge}$/$L_{\odot})^{1.02}$ and \ngc\ $\propto$
$(M_{*,bulge}$/$M_{\odot})^{0.87}$.  The fit of log(\ngc) vs.\ bulge
luminosity yields $\chi^2$/$(N-2)$ $=$ 9.2 and $\epsilon_0$ $=$
0.17~dex, whereas the linear fit of log(\ngc) vs.\ bulge mass gives
$\chi^2$/$(N-2)$ $=$ 11.5 and $\epsilon_0$ $=$ 0.23~dex.  By these
measures, the tightest correlation of log(\ngc) with any of the galaxy
properties examined here is still the correlation with
galaxy stellar mass, shown in the right-hand panel of Figure~\ref{fig:ngc lum
  galmass} and given in Equation~\ref{equation:bf ngc galmass}. 

One might assume that if it is the red, metal-rich GCs that are
associated with a galaxy's spheroidal component, then perhaps the
numbers of red GCs will better correlate with bulge luminosity or
mass. Including only the red GCs in log(\ngc) and fitting the log of
the bulge luminosity or mass produces linear relations with the same
slope (within the errors) as those in Equations~\ref{equation:bf ngc
  lum bulge} and \ref{equation:bf ngc galmass bulge}, but larger
$\chi^2$ and $\epsilon_0$ values.  In fact, I also tried executing the
fits counting only the blue GCs in each galaxy and found that counting
the blue GCs produces a better-quality fit than counting only the red
ones.  Still, using the total number of GCs produces the best results
in terms of $\chi^2$ and $\epsilon_0$.

To summarize, the relationship between \ngc\ and galaxy mass, or even
with bulge luminosity or mass, for the 30 galaxies in this sample
yields reduced $\chi^2$ values that are smaller than or comparable to
the smallest reduced $\chi^2$ values yielded by the \ngc$-$\mbh\ fits
in Section~\ref{section:ngc mbh}.  Furthermore, the intrinsic scatter
is much smaller for the fits explored in this section ($\epsilon_0$
$\sim$ 0.1$-$0.2~dex) than for either the \ngc$-$\mbh\ or the
M$-$$\sigma$ relations 
(both of which have $\epsilon_0$ \gapp 0.4 dex).  This seems to imply
that the
relationship between \ngc\ and the luminosity or mass of the host
galaxy 
is the more fundamental one, and that the observed correlation between
\ngc\ and SMBH mass results from both of these quantities being
correlated with galaxy mass.
As noted earlier, Snyder et al.\ (2011) came to a similar conclusion
based on their analysis of the residuals in the relation between
\ngc\ and \mbh\ for a sample of elliptical galaxies drawn from BT10
and HH11.  The result here is therefore not unexpected since --- as
the previous papers on this topic (BT10, HH11, and Snyder et
al.\ 2011) have all noted --- it seems highly unlikely that the GCs in
a galaxy will have some sort of direct causal link to the SMBH located
in the galaxy center.

It is relevant to discuss here a recent paper by Sadoun \& Colin
(2012) that examined other correlations between GC systems and SMBH
masses.  For a sample of twelve giant spiral, S0, and elliptical
galaxies, Sadoun \& Colin (2012) find a correlation between the
projected velocity dispersion of the GC system ($\sigma_{\rm GC}$) and
the SBMH mass that is as tight as the $M$-$\sigma$ relation for those
twelve galaxies.  In the discussion section of their paper, Sadoun \&
Colin point out that their observational results might be understood
in the context of recent numerical simulations of galaxy formation
done by Jahnke \& Maccio (2011).  Jahnke \& Maccio simulate the
hierarchical assembly and merger history of dark matter halos over
cosmic time, with added recipes for star formation, black hole
accretion, and bulge evolution in the simulated galaxies.  They then
examine the scaling relation between \mbh\ and bulge mass that is
produced in the resultant galaxies at $z$$=$0.  Their conclusion is
that correlations between SMBH mass and bulge mass (or velocity
dispersion as a proxy for mass) do not require a direct physical link,
but rather can arise through normal galaxy merging in a $\Lambda$CDM
universe that occurs over the course of a giant galaxy's history, from
high-redshift until $z$$=$0.  In Jahnke \& Maccio's simulations, the
scaling relations are a natural outcome of the simultaneous build-up
of both the central black hole and the galaxy's stellar component by
major galaxy mergers.  Jahnke \& Maccio state that, ``we in principle
expect a correlation with \mbh\ for any (mass) parameter that is
subject to the same $\Lambda$CDM assembly chain'', including
correlations between GC populations and \mbh.  Sadoun \& Colin (2012)
point out that this could provide an explanation for their observed
\mbh\ and $\sigma_{\rm GC}$ results, and likewise it seems to provide
a broad context within which we might understand the correlations
between the various parameters presented in the current paper.

\subsection{Deriving an $M$-$\sigma$ Relation for the Wide-Field GC System Survey
  Galaxy Sample}
\protect\label{section:wfgcss m sigma}

One last piece of analysis that can be done with the data set compiled
here is to calculate predicted SMBH masses for the giant spiral and
elliptical galaxies in Table~\ref{table:wfgcss}.  These ten galaxies
have well-measured \ngc\ values from my wide-field survey but do not
yet have measured masses in the literature for the SMBH that
presumably exists in each of their
central regions.  Nine of the ten galaxies in Table~\ref{table:wfgcss}
are spiral galaxies and one (NGC~4406) is a massive Virgo Cluster
elliptical.  Consequently it seems appropriate to derive \mbh\ from
the relationship between \ngc\ and \mbh\ given in
Equation~\ref{equation:all bf}, i.e., the linear relation based on the
full sample of elliptical, S0, and spiral galaxies analyzed in
Section~\ref{section:ngc mbh} and listed in Table~\ref{table:master},
rather than using the relation derived from only the E/S0 galaxies
(Equation~\ref{equation:ES0 bf}).  The calculated SMBH masses 
are listed in column (2) of Table~\ref{table:wfgcss}.  The errors on
the masses are computed by determining the predicted SMBH mass for the
upper and lower end of the possible \ngc\ values.  For example,
\ngc\ for the spiral galaxy NGC~1055 is 220$\pm$30, so the
uncertainties in \mbh\ reflect the possible range of SMBH values if
\ngc\ is 220$-$30$=$190 or 220$+$30$=$250.
Published measurements of the central velocity dispersion for nine of
the ten galaxies are given in the third column of the table. One
galaxy (NGC~7339) apparently does not have a published velocity
dispersion, so its SMBH mass is listed in the table but it is excluded
from the rest of the analysis in this section.

\begin{figure}
\includegraphics[width=3.5in]{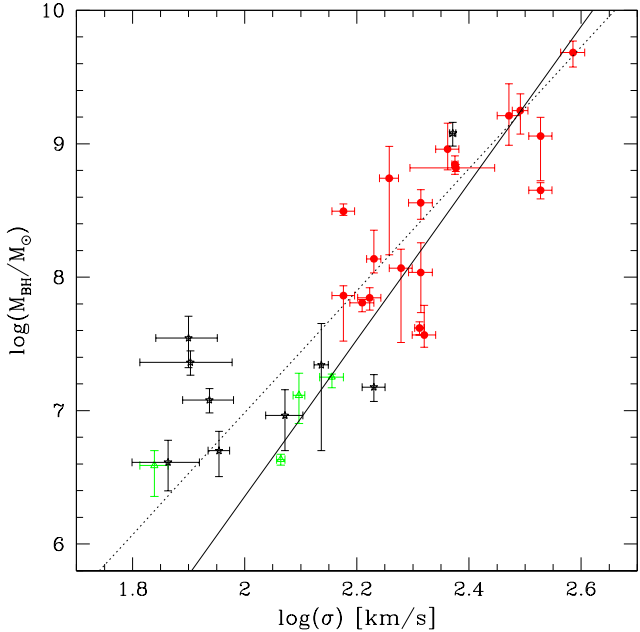}
\caption{\normalsize The 
log of the measured or predicted SMBH mass versus the log of the
central velocity dispersion for the 30 giant galaxies in
Tables~\ref{table:master} and \ref{table:wfgcss}. As in previous
figures, galaxies from Table~\ref{table:master} are plotted
with red filled circles (classical bulges) or green open triangles
(pseudobulges) and the \mbh\ data plotted are {\it measured} values.
Open stars denote galaxies from Table~\ref{table:wfgcss} and the
\mbh\ values are {\it predicted} based on the number of GCs in the
galaxy and Equation~\ref{equation:all bf}. The lines are the same as
those plotted in the earlier version of this figure, Figure~\ref{fig:m
  sigma}: the solid line is the best-fitting line for the 20 galaxies
in Table~\ref{table:master} and the dotted line is the $M$-$\sigma$
relation from BT10.
}  
\protect\label{fig:m sigma wfgcss}
\end{figure}

Figure~\ref{fig:m sigma wfgcss} shows the nine galaxies from
Table~\ref{table:wfgcss} with predicted SMBH masses and measured
velocity dispersions, added to the $M-\sigma$ plot shown earlier in
this paper (Figure~\ref{fig:m sigma}).  As before, the red filled
circles are classical-bulge galaxies and the green open triangles are
pseudobulge galaxies.  The open stars are the nine new galaxies from
the wide-field GC system survey.  The two lines are the same as those
shown in the first $M-\sigma$ plot: the solid line is the best-fitting
$M-\sigma$ relation given in Equation~\ref{equation:m-sigma all} and the
dotted line is the best-fitting $M-\sigma$ relation published in BT10.

Most of the galaxies from the wide-field survey do seem to follow the
general relationship between \mbh\ and velocity dispersion shown by
the two lines. If one adds the data points for these nine galaxies to
the data points from Table~\ref{table:master} and fits a linear
relation to the 29 points, the result is a line with the same slope
and intercept (5.88$\pm$0.19 and 8.19$\pm$0.03, respectively) within
the errors as those of Equation~\ref{equation:m-sigma all}, and
similar reduced $\chi^2$ values ($\chi^2$/$(N-4)$ $=$ 12.5 and
$\chi^2$/$(N-2)$ $=$ 11.6).  On the other hand, a few spiral galaxies
from the survey --- namely, NGC~1055, NGC~3556, and to a lesser extent
NGC~4013 --- deviate noticeably from the relations in that they have
very large \mbh\ values for their measured velocity dispersions.  One
relevant question to ask is: do these galaxies have large predicted
SMBH masses because they have anomalously large GC populations?  The
answer is no: all three of these galaxies have the expected number of
GCs based on their stellar mass, i.e., they are not outliers on the
\ngc\ versus mass plot (see Figure~\ref{fig:ngc lum galmass} and the
data in Table~\ref{table:wfgcss}).  What the data seem to be
indicating instead is that \ngc\ is not a particularly good predictor
of SMBH mass for some galaxies.  The three galaxies that deviate most
strongly in the $M-\sigma$ plot are all spiral galaxies.  HH11 pointed
out that the total number of GCs in the Milky Way was also not a good
predictor of its SMBH mass; that is, it deviated strongly from their
best-fitting \ngc$-$\mbh\ relation and Figure~\ref{fig:ngc smbh} shows
that it deviates from the relation derived here.
%

The one additional elliptical galaxy from the wide-field survey is
NGC~4406, which appears at $log(M_{\rm BH}/M_{\odot})$ $\sim$ 9.1 and
$log(\sigma)$ $\sim$ 2.4 in Figure~\ref{fig:m sigma wfgcss}.  The
spiral galaxies all have relatively low \mbh\ and velocity dispersion
values and lie in the lower left region of the figure.  If one
considers only the spiral galaxies from the survey, it appears that
they actually do not seem to follow an $M-\sigma$ relation and it is
really the data point for NGC~4406 that is driving the agreement
between the new data and the original best-fitting $M-\sigma$ line.
In fact, excluding NGC~4406 from the sample and fitting a line to the
other eight data points from the wide-field survey yields a line with
zero slope within the errors:

\begin{equation}
{\rm{log\frac{M_{\rm SMBH}}{M_{\odot}} = (7.19\pm0.23) + (0.17\pm0.80)
  log\frac{\sigma}{200~km~s^{-1}}}}
\label{equation:m-sigma eight spirals}
\end{equation}

\noindent Although it is not entirely obvious what this means, what it
seems to suggest is that, again, \ngc\ may not be an accurate
predictor of SMBH mass, at least not for spiral galaxies; if there
were a tight correlation between the two quantities, one might expect
the spiral galaxies to follow the expected $M-\sigma$ relation more
closely.  A larger sample of galaxies with well-determined values of
\ngc, \mbh, bulge velocity dispersion, and bulge classification would
presumably help to
clarify these issues.

\section{Summary}

I have assembled a sample of 20 giant galaxies with both measured SMBH
masses and reliable estimates of \ngc\ and GC color fractions based on
high-quality wide-field imaging data.  The sample includes eight
galaxies from my ongoing GC system survey, four galaxies from the
ACSVCS survey (Peng et al.\ 2008), two galaxies from a Gemini study by
Faifer et al.\ (2011), and six galaxies from various studies in the
literature.  Half of the galaxies in the sample are ellipticals, eight
are S0 galaxies, and two are spiral galaxies (the Milky Way and M31).
Four of the galaxies may possess pseudobulges according to previous
studies of their light profiles and/or kinematics.  The sample of 20
galaxies is used to explore correlations between the galaxies' GC
populations and the masses of their central SMBHs. The main findings
of this investigation are as follows:

\begin{enumerate}
\item The E/S0 galaxies in the sample follow a relation between the log of
\ngc\ and the log of the SMBH mass of the form:

\begin{equation}
\rm{{log\frac{M_{\rm SMBH}}{M_{\odot}} = (8.04\pm0.03) + (1.22\pm0.06)
  log\frac{N_{GC}}{500}}}
\end{equation}

\noindent When the two spiral galaxies are included in the sample, the
slope and intercept of the relation become 1.52$\pm$0.06 and
7.91$\pm$0.03, respectively. The coefficients of the best-fitting
relation for E/S0 galaxies agree within the errors with those of the
corresponding relation for early-type galaxies from BT10 (who found a
slope of 1.08$\pm$0.04 and an intercept of 8.14$\pm$0.04) and the
relation for elliptical galaxies from HH11 (who derived a slope of
0.98$\pm$0.10 and an intercept of 8.30$\pm$0.29).

\item The $M$-$\sigma$ relation for the 20 galaxies in the sample
  agrees with those in the recent literature within the errors.  Three
  of the four galaxies with pseudobulges (including the Milky Way)
  closely follow the best-fitting linear relation defined by the
  classical-bulge galaxies.

\item The numbers of blue, metal-poor GCs in the galaxies yield a {\it
  slightly} smaller reduced $\chi^2$ than the numbers of red,
  metal-rich GCs, although this small difference is probably not
  statistically significant.
In the current picture of galaxy and GC system formation, metal-poor
GCs originate in the earliest epoch of galaxy assembly.  If metal-poor
GC populations were actually more tightly correlated to the SMBH masses
in the host galaxies, this might imply that the correlation between
\ngc\ and \mbh\ is put in place early in the history of the galaxy and
does not depend strongly on the occurrence of major merger events
later.

\item When the galaxy sample is divided according to Hubble type, the
  elliptical galaxies retain the tight correlation between \ngc\ and
  SMBH mass seen in the full sample, whereas the S0 galaxies show more
  scatter.  This is similar to the result in HH11, who found that the
  S0 galaxies show a large dispersion in black hole mass for a given
  \ngc.  The four galaxies that may have pseudobulges also seem to
  show increased scatter compared to the elliptical and
  classical-bulge galaxies.  However, the number of pseudobulge
  galaxies in the sample is small so it is not clear whether or not
  the latter trend is real and therefore implies something about how
  SMBHs grow.

\item Ten more galaxies from my wide-field GC system survey were used
  to supplement the original sample and explore correlations between
  \ngc\ and other galaxy properties.  The strongest correlation with
  the smallest intrinsic scatter is a correlation between the number
  of GCs and the total stellar mass of the host galaxy.  In general
  the correlations between \ngc\ and galaxy mass and luminosity or
  bulge mass and luminosity are much tighter than any of the
  \ngc$-$\mbh\ correlations for this sample.  This seems to confirm
  the idea that the observed connection between \ngc\ and \mbh\ in
  giant galaxies is a consequence of the connection between both of
  these quantities and the galaxy potential.

\item Finally, the \ngc$-$\mbh\ relation derived here is used to
  calculate predicted SMBH masses for the ten additional galaxies with
  measured \ngc\ from the wide-field survey but without existing
  \mbh\ measurements in the literature.  The single elliptical galaxy
  in this subsample lies close to the expected $M-\sigma$ relation,
  but the spiral galaxies show larger scatter, suggesting that \ngc\ is
  not a reliable predictor of SMBH mass for some galaxies.\\
\end{enumerate}


\acknowledgments 
The research described in this paper was supported by
an NSF Faculty Early Career Development (CAREER) award (AST-0847109).
I thank an anonymous referee for carefully reading the manuscript and
providing valuable suggestions that improved the paper.  I also thank
Will Clarkson, Tom Maccarone, and Samir Salim for useful discussions
regarding various aspects of this work, and Karl Gebhardt for
providing an estimate of the SMBH mass of NGC~4472 from a paper in
preparation.  Jonathan Hargis assisted me with the initial
implementation of the {\tt fitexy} algorithm.  I thank Gregory Snyder
and Scott Tremaine for promptly answering my detailed questions about
their papers.  Finally, I am grateful to the staff of the WIYN
Observatory and Kitt Peak National Observatory for their assistance
with obtaining some of the observational data that contributed to this
study.  This research has made use of the NASA/IPAC Extragalactic
Database (NED) which is operated by the Jet Populsion Laboratory,
California Institute of Technology, under contract with the National
Aeronautics and Space Administration.




\begin{deluxetable}{lcccclcccclr}
\tabletypesize{\scriptsize}
\tablecaption{Giant Galaxies with Measured SMBH Masses and Global GC System Properties}
\tablewidth{0pt}
\tablehead{
Name & Type & $M_{BH}$ & Ref & $\sigma$ & $M_V$ & Dist & \ngc\ & $S_N$ & $T$ & $f_{\rm blue}$ & Ref\\
  &   & ($M_{\odot}$) & & (km/s) &  & (Mpc) &  & & \\ 
}
\startdata 
N821 & E6  & 3.7($+$2.5,$-$0.7)x10$^7$ & 12 & 209$\pm$10\tablenotemark{a} &  -21.0 & 22.4 &  320$\pm$45 &  1.3$\pm$0.2 &  1.5$\pm$0.2  & 0.70 & 22\\
N1399\tablenotemark{\dag} & E1pec  & 1.1($+$0.4,$-$0.6)x10$^9$ & 12 & 337$\pm$16\tablenotemark{a} &  -22.3 & 18.5 & 5800$\pm$700\tablenotemark{e} &  7.0$\pm$0.8 &  8.2$\pm$1.0  & 0.60 & 3,5\\
  &   & 4.5($+$0.6,$-$0.6)x10$^8$ & & & & & &\\
N3379\tablenotemark{\dag} & E1  & 1.1($+$0.7,$-$0.5)x10$^8$ & 7,12 & 206$\pm$10\tablenotemark{a} &  -20.9 & 10.6 &  274$\pm$31 &  1.2$\pm$0.3 &  1.4$\pm$0.3  & 0.70 & 20\\
  &   & 3.6($+$0.9,$-$0.9)x10$^8$ & & & & & &\\
N4374 & E1  & 1.6($+$1.2,$-$0.6)x10$^9$ & 12 & 296$\pm$14\tablenotemark{a} &  -22.1 & 18.4 & 1775$\pm$381 &  2.7$\pm$0.6 &  3.1$\pm$0.7  & 0.59 & 10\\
N4472 & E2  & 1.8($+$0.6,$-$0.6)x10$^9$ & 8 & 310$\pm$10\tablenotemark{b} &  -23.1 & 16.7 & 5870$\pm$680 &  3.6$\pm$0.6 &  4.2$\pm$0.6  & 0.60 & 19\\
N4473 & E5  & 1.2($+$0.4,$-$0.8)x10$^8$ & 12 & 190$\pm$9\tablenotemark{a} &  -20.7 & 15.3 &  376$\pm$97 &  2.0$\pm$0.5 &  2.3$\pm$0.6  & 0.57 & 17\\
N4564 & E6  & 6.4($+$0.4,$-$0.9)x10$^7$ & 12 & 162$\pm$8\tablenotemark{a} &  -19.9 & 15.8 &  213$\pm$31 &  2.2$\pm$0.3 &  2.6$\pm$0.4  & 0.62 & 17\\
N4649 & E2  & 4.8($+$1.1,$-$1.1)x10$^9$ & 21 & 385$\pm$19\tablenotemark{a} &  -22.4 & 16.8 & 4073$\pm$820 &  4.5$\pm$0.9 &  5.2$\pm$1.1  & 0.52 & 6\\
N5128\tablenotemark{\dag} & Epec  & 3.1($+$0.4,$-$0.2)x10$^8$ & 12 & 150$\pm$7\tablenotemark{a} &  -22.0 &  4.6 &  980$\pm$120 &  1.5$\pm$0.2 &  1.8$\pm$0.2  & 0.53 & 15\\
  &   & 7.3($+$1.4,$-$3.9)x10$^7$ & & & & & &\\
N5813 & E1-2  & 7.0($+$1.1,$-$1.1)x10$^8$ & 11 & 237$\pm$3\tablenotemark{b} &  -22.3 & 32.2 & 3100$\pm$800 &  3.9$\pm$0.9 &  4.5$\pm$1.1  & 0.57 & 13\\
N1023 & SB0  & 4.2($+$0.5,$-$0.5)x10$^7$ & 11 & 205$\pm$4\tablenotemark{b} &  -21.1 & 11.4 &  494$\pm$80 &  1.7$\pm$0.3 &  2.7$\pm$0.4  & 0.58 & 23\\
N3115 & S0  & 9.1($+$5.1,$-$2.8)x10$^8$ & 12 & 230$\pm$11\tablenotemark{a} &  -21.1 &  9.7 &  439$\pm$130 &  1.6$\pm$0.3 &  2.4$\pm$0.7  & 0.52 & 6\\
N3384 & SB0  & 1.8($+$0.1,$-$0.3)x10$^7$ & 12 & 143$\pm$7\tablenotemark{a} &  -20.5 & 11.6 &  122$\pm$30 &  0.8$\pm$0.2 &  1.2$\pm$0.4  & 0.60 & 13\\
N4350 & S0  & 5.5($+$4.0,$-$4.0)x10$^8$ & 11 & 181$\pm$7\tablenotemark{b} &  -20.1 & 16.5 &  196$\pm$60 &  1.8$\pm$0.5 &  2.7$\pm$0.8  & 0.71 & 17\\
N4459 & S0  & 7.0($+$1.3,$-$1.3)x10$^7$ & 12 & 167$\pm$8\tablenotemark{a} &  -20.9 & 16.1 &  218$\pm$28 &  1.0$\pm$0.1 &  1.5$\pm$0.2  & 0.52 & 17\\
N4594 & S0/Sa  & 6.6($+$0.4,$-$0.4)x10$^8$ & 16 & 238$\pm$41\tablenotemark{c} &  -22.4 &  9.8 & 1890$\pm$200 &  2.1$\pm$0.3 &  3.2$\pm$0.5  & 0.63 & 20\\
N7332 & S0pec  & 1.3($+$0.6,$-$0.5)x10$^7$ & 12 & 125$\pm$3\tablenotemark{b} &  -20.8 & 23.0 &  175$\pm$56 &  0.9$\pm$0.3 &  1.4$\pm$0.4  & 0.57 & 23\\
N7457 & S0  & 3.9($+$1.1,$-$1.6)x10$^6$ & 12 &  69$\pm$4\tablenotemark{b} &  -19.5 & 13.2 &  210$\pm$30 &  3.1$\pm$0.7 &  4.8$\pm$1.1  & 0.58 & 14\\
Milky Way & Sbc  & 4.3($+$0.4,$-$0.4)x10$^6$ & 9 & 116$\pm$2\tablenotemark{d} &  -21.3 & 0.0083 &  160$\pm$20 &  0.5$\pm$0.1 &  1.1$\pm$0.1  & 0.70 & 1\\
M31 & Sb  & 1.4($+$0.9,$-$0.3)x10$^8$ & 4 & 170$\pm$5\tablenotemark{b} &  -21.8 &  0.8 &  450$\pm$100 &  0.9$\pm$0.2 &  1.6$\pm$0.4  & 0.72 & 1,2,18\\

\enddata 
\tabletypesize{\scriptsize}
\tablecomments{Morphological types are from the NASA/IPAC
  Extragalactic Database (NED), with three exceptions. NED lists
  NGC~5128 as an ``S0 pec'' or ``E/S0'', but studies of this galaxy's
  properties and stellar populations (e.g., Hui et al. 1993, Peng et
  al.\ 2004, Harris et al.\ 2004) describe it as an elliptical.
  NGC~4594 is often classified as an Sa but it has a bulge fraction of
  $B/T$$=$0.86 (Kent 1988) and broadband colors like an S0, so I list
  it here as ``S0/Sa''. The Milky Way is listed as an Sbc, which is
  the classification commonly used in the literature (e.g., Ashman \&
  Zepf 1998).}
\tablenotetext{\dag}{Two estimates of SMBH mass are
  given for the galaxies NGC~1399 and NGC~5128 in Gultekin et
  al.\ (2009), as well as for NGC~3379 in BT10.  I adopt the approach
  taken by Gultekin et al.\ (2009) and BT10 and use both SMBH mass
  measurements for these galaxies, but increase the error bars by a
  factor of $\sqrt2$, to give each measurement half-weight.}
\tablenotetext{a}{Velocity dispersion from Gultekin et al.\ (2009).}
\tablenotetext{b}{Velocity dispersion from HYPERLEDA.}
\tablenotetext{c}{Velocity dispersion from Jardel et al.\ (2011).}
\tablenotetext{d}{Velocity dispersion from Howard et al.\ (2008).}
\tablenotetext{e}{The GC system properties of NGC~1399 were calculated
  by Spitler et al.\ (2008), using measurements drawn from both Dirsch
  et al.\ (2003) and Bassino et al.\ (2006).}
\tablerefs{(1) Ashman \& Zepf 1998; (2) Barmby et al.\ 2000; (3)
  Bassino et al.\ (2006); (4) Bender et al.\ 2005; (5) Dirsch et
  al.\ 2003; (6) Faifer et al.\ 2011; (7) Gebhardt et al.\ 2000a; (8)
  Gebhardt et al.\ 2012, in preparation; (9) Gillessen et al.\ 2009;
  (10) Gomez \& Richtler 2004; (11) Graham 2008; (12) Gultekin et
  al.\ 2009; (13) Hargis \& Rhode 2012; (14) Hargis et al.\ 2011; (15)
  Harris et al.\ 2004; (16) Jardel et al.\ 2011; (17) Peng et
  al.\ 2008; (18) Perrett et al.\ (2002); (19) Rhode \& Zepf 2001;
  (20) Rhode \& Zepf 2004; (21) Shen \& Gebhardt 2010; (22) Spitler et
  al.\ 2008; (23) Young, Dowell, \& Rhode 2012.}
\protect\label{table:master}
\end{deluxetable}
%

\begin{deluxetable}{lrccccl}
\tablecaption{Results of Fitting Process for \ngc\ and \mbh\ Data} 
\tablewidth{0pt}
\tablehead{
Sample & $N_{\rm points}$ & $\alpha$ & $\beta$ & $\chi^2$/$(N-4)$ & $\chi^2$/$(N-2)$ & $\epsilon_0$
}
\startdata 
E/S0s & 21 & 8.04$\pm$0.03 & 1.22$\pm$0.06 &  10.4 & 9.3 & 0.45 \\
All & 23 & 7.91$\pm$0.03 & 1.52$\pm$0.06 & 12.4 & 11.2 & 0.50 \\
E/S0s, blue GCs & 21 & 8.32$\pm$0.03 & 1.23$\pm$0.06 & 10.6 & 9.5 & 0.48\\
E/S0s, red GCs  & 21 & 8.52$\pm$0.03 & 1.20$\pm$0.05 & 11.0 & 9.8 & 0.51\\
Es & 13 & 8.17$\pm$0.04 & 0.95$\pm$0.07 & 9.4 & 7.7 & 0.37\\
S0s (with N4594) & 8 & 7.98$\pm$0.05 & 1.63$\pm$0.14 & 16.2 & 10.8& 0.70\\
S0s (without N4594) & 7 & 10.62$\pm$2.52 & 10.21$\pm$5.46 & 9.8 & 5.9 & 0.75\\
\enddata \tablecomments{For the fits between \ngc\ and \mbh, the form
  of the linear relation is  
log$\frac{M_{\rm SMBH}}{M_{\odot}}$ = $\alpha$ + $\beta$log$\frac{N_{GC}}{500}$.
Values of $\epsilon_0$ in column (7) are
  derived by adding $\epsilon_0$ in quadrature to the errors on the
  data points.}
\protect\label{table:fitting results}
\end{deluxetable}

\begin{deluxetable}{lccclcccclr}
\tabletypesize{\scriptsize}
\tablecaption{Giant Galaxies with Measured GC System Properties from the Wide-Field Survey and Predicted SMBH Masses}
\tablewidth{0pt}
\tablehead{
Name & Type & Predicted $M_{BH}$ & $\sigma$ & $M_V$ & Distance & \ngc\ & $S_N$ & $T$ & $f_{\rm blue}$ & Ref\\
  &   & ($M_{\odot}$) & (km/s) &  & (Mpc) &  & & & \\ 
}
\startdata 
N4406 & E3  & 1.2($+$0.2,$-$0.2)x10$^9$ & 235$\pm$3\tablenotemark{b} &  -22.3 & 16.7 & 2900$\pm$400 &  3.5$\pm$0.5 &  4.1$\pm$0.6  & 0.62 & 3\\
N891 & Sb  & 4.1($+$1.9,$-$1.6)x10$^6$ &  73$\pm$10\tablenotemark{b} &  -20.8 &  8.4 &   70$\pm$20 &  0.3$\pm$0.1 &  0.6$\pm$0.3  & 0.50 & 1\\
N1055 & Sb  & 2.3($+$0.5,$-$0.5)x10$^7$ &  80$\pm$15\tablenotemark{b} &  -21.0 & 16.3 &  220$\pm$30 &  0.9$\pm$0.2 &  1.7$\pm$0.5  & 0.59 & 5\\
N2683 & Sb  & 9.2($+$5.1,$-$4.2)x10$^6$ & 118$\pm$9\tablenotemark{b} &  -20.5 &  7.7 &  120$\pm$40 &  0.8$\pm$0.4 &  1.4$\pm$0.7  & 0.68 & 4\\
N3556 & Sc  & 3.5($+$1.6,$-$1.4)x10$^7$ &  79$\pm$10\tablenotemark{a} &  -21.2 & 12.4 &  290$\pm$80 &  0.9$\pm$0.4 &  2.2$\pm$0.9  & 0.64 & 4\\
N4013 & Sb  & 1.2($+$0.3,$-$0.2)x10$^7$ &  86$\pm$9\tablenotemark{a} &  -20.4 & 15.1 &  140$\pm$20 &  1.0$\pm$0.2 &  1.9$\pm$0.5  & 0.74 & 1\\
N4157 & Sb  & 5.0($+$2.0,$-$1.8)x10$^6$ &  90$\pm$4\tablenotemark{b} &  -20.4 & 14.7 &   80$\pm$20 &  0.6$\pm$0.3 &  1.1$\pm$0.6  & 0.55 & 4\\
N7331 & Sb  & 2.2($+$2.3,$-$1.7)x10$^7$ & 137$\pm$4\tablenotemark{b} &  -21.7 & 13.1 &  210$\pm$130 &  0.5$\pm$0.3 &  0.8$\pm$0.7  & 0.50 & 4\\
N7339 & Sbc  & 4.5($+$0.9,$-$0.9)x10$^6$ & ... &  -20.4 & 22.4 &   75$\pm$10 &  0.5$\pm$0.2 &  1.5$\pm$0.5  & 0.60 & 5\\
N7814 & Sab  & 1.5($+$0.4,$-$0.3)x10$^7$ & 170$\pm$8\tablenotemark{b} &  -20.4 & 13.2 &  165$\pm$25 &  1.3$\pm$0.4 &  2.2$\pm$0.8  & 0.58 & 2\\

\enddata \tablecomments{Morphological
  types are from the NASA/IPAC Extragalactic Database (NED).}
\tablenotetext{a}{Velocity dispersion from Ho et al.\ (2009).}
\tablenotetext{b}{Velocity dispersion from HYPERLEDA.}
\tablerefs{(1) Rhode, Windschitl, \& Young 2010; (2) Rhode \& Zepf
  2003; (3) Rhode \& Zepf 2004; (4) Rhode et al.\ 2007; (5) Young,
  Dowell, \& Rhode 2012.}
\protect\label{table:wfgcss}
\end{deluxetable}


\begin{thebibliography}{}

\bibitem[Alves-Brito et al.\ (2011)]{alves11} Alves-Brito, A., Hau,
  G.K.T., Forbes, D.A., Spitler, L.R., Strader, J., Brodie, J.P., \&
  Rhode, K.L. 2011, \mnras, 417, 1823

\bibitem[Andrae et al.\ (2010)]{andrae10b} Andrae, R., Schulze-Hartung,
  T., \& Melchior, P. 2010, arXiv:1012.3754

\bibitem[Ashman \& Zepf (1992)]{az92} Ashman, K.M., \& Zepf, S.E. 1992,
\apj, 384, 50

\bibitem[Ashman \& Zepf (1998)]{az98} Ashman, K.M., \& Zepf, S.E. 1998,
Globular Cluster Systems (Cambridge: Cambridge University Press)

\bibitem[Barmby et al.\ (2000)]{barmby00} Barmby, P., Huchra, J.P.,
  Brodie, J.P., Forbes, D.A., Schroder, L.L., \& Grillmair, C.J. 2000,
  \aj, 119, 727

\bibitem[Barmby et al.\ (2001)]{barmby01} Barmby, P., Huchra, J.P.,
  Brodie, J.P. 2001, \aj, 121, 1482

\bibitem[Baum (1955)]{baum55} Baum, W.A. 1955, \pasp, 67, 328

\bibitem[Bassino et al.\ (2006)]{bass06} Bassino, L.P., Faifer, F.R.,
  Forte, J.C., Dirsch, B., Richtler, T., Geisler, D., \& Schuberth,
  Y. 2006, A\&A, 451, 789

\bibitem[Bastian et al.\ (2006)]{bastian06} Bastian, N., Saglia, R.P.,
  Goudfrooij, P., Kissler-Patig, M., Maraston, C., Schweizer, F., \&
  Zoccali, M. 2006, \aa, 448, 881

\bibitem[Beasley et al.\ (2002)]{beasley02} Beasley, M.A., Baugh,
C.M., Forbes, D.A., Sharples, R.M., \& Frenk, C.S. 2002, \mnras, 333, 383

\bibitem[Beasley et al.\ (2008)]{beasley08} Beasley, M.A., Bridges,
  T., Peng, E., Harris, W.E., Harris, G.L.H., Forbes, D.A., \& Mackie,
  G. 2008, \mnras, 386, 1443

\bibitem[Bender et al.\ (2005)]{bender05} Bender, R. et al.\ 2005,
  \apj, 631, 280

\bibitem[Binney \& Merrifield (1998)]{bm98} Binney, J. \& Merrifield,
  M. 1998, Galactic Astronomy (Princeton: Princeton Univ.\ Press),
  p.\ 220

\bibitem[Brodie \& Strader (2006)]{brodie06} Brodie, J.P. \& Strader,
  J. 2006, \araa, 44, 193

\bibitem[Burkert \& Silk (2001)]{bs01} Burkert, A. \& Silk, J. 2001,
  \apj, 554, 151

\bibitem[Burkert \& Tremaine (2010)]{bt10} Burkert, A. \& Tremaine,
  S. 2010, \apj, 720, 516

\bibitem[C\^ot\'e et al.\ (1998)]{cote98} C\^ot\'e, P., Marzke, R.O., \& West,
M.J. 1998, \apj, 501, 554

\bibitem[C\^ot\'e et al.\ (2000)]{cote00} C\^ot\'e, P., Marzke, R.O.,
  \& West, M.J., \& Minniti, D. 2000, \apj, 533, 869

\bibitem[C\^ot\'e et al.\ (2004)]{cote04} C\^ot\'e, P., Blakeslee,
  J.P., Ferrarese, L., Jordan, A., Mei, S., Merritt, D.,
  Milosavijevic, M., Peng, E.W., Tonry, J.L., \& West, M.J. 2004,
  ApJS, 153, 223

\bibitem[Dawe \& Dickens (1976)]{dd76} Dawe, J.A. \& Dickens,
  R.J. 1976, Nature, 263, 395

\bibitem[de Jong et al.\ (2008)]{dejong08} de~Jong, R. S.,
  Radburn-Smith, D. J., \& Sick, J. N. 2009, in IAU Symp. 254, The
  Galaxy Disk in Cosmological Context, ed. J. Andersen,
  J. Bland-Hawthorn, \& B. Nordstrom (Cambridge: Cambridge
  Univ. Press), 19


\bibitem[de Vaucouleurs et al.\ (1991)]{devauc91} de~Vaucouleurs, G.,
de~Vaucouleurs, A., Corwin Jr., H.G., Buta, R.J., Paturel, G., \&
Fouque, P. 1991, Third Reference Catalogue of Bright Galaxies (New
York: Springer)

\bibitem[Di Matteo et al.\ (2005)]{dimatteo05} Di~Matteo, T.,
  Springel, V., \& Hernquist, L. 2005, Nature, 433, 604

\bibitem[Dirsch et al.\ (2003)]{dirsch03} Dirsch, B., Richtler, T.,
  Geisler, D., \& Forte, J.C., Bassino, L.P., \& Gieren, W.P. 2003,
  \aj, 125, 1908

\bibitem[Dotter et al.\ (2010)]{dotter10} Dotter, A. et al.\ 2010,
  \apj, 708, 698

\bibitem[Faber \& Jackson (1976)]{faber76} Faber, S.M. \& Jackson,
  R.E. 1976, \apj, 204, 668

\bibitem[Faifer et al.\ (2011)]{faifer11} Faifer, F.R., Forte, J.C.,
  Norris, M.A., Bridges, T., Forbes, D.A., Zepf, S.E., Beasley, M.,
  Gebhardt, K., Hanes, D.A., \& Sharples, R.M. 2011, \mnras, 416, 155

\bibitem[Falcon-Barroso et al.\ (2004)]{falcon04} Falcon-Barroso, J.,
  Peletier, R. F., Emsellem, E., Kuntschner, H., Fathi, K., Bureau,
  M., Bacon, R., Cappellari, M., Copin, Y., Davies, R. L., de Zeeuw,
  T. 2004, \mnras, 350, 35

\bibitem[Fan et al.\ (2006)]{fan06} Fan, X. et al.\ 2006, \aj, 131,
  1203 

\bibitem[Ferrarese \& Ford (2005)]{ferrarese05} Ferrarese, L. \& Ford,
  H. 2005, SSRv, 116, 523

\bibitem[Ferrarese \& Merritt (2000)]{ferr00} Ferrarese, L. \&
  Merritt, D. 2000, \apj, 539, L9

\bibitem[Fisher \& Drory (2008)]{fd08} Fisher, D.B. \& Drory, N. 2008,
  \aj, 136, 773

\bibitem[Fisher \& Drory (2010)]{fd10} Fisher, D.B. \& Drory, N. 2010,
  \apj, 716, 942

\bibitem[Forbes \& Bridges (2010)]{fb10} Forbes, D.A. \& Bridges,
  T. 2010, \mnras, 404, 1203

\bibitem[Forbes et al.\ (1997)]{fbg97} Forbes, D.A., Brodie, J.P., \&
Grillmair, C.J. 1997, \aj, 113, 1652

\bibitem[Forbes et al.\ (2001)]{fbl01} Forbes, D.A., Brodie, J.P., \&
  Larsen, S.S. 2001, \apj, 556, L83

\bibitem[Gebhardt et al.\ (2000a)]{gebhardt00a} Gebhardt, K., Richstone,
  D., Kormendy, J., Lauer, T.R., Ajhar, E.A., Bender, R., Dressler,
  A., Faber, S.M., Grillmair, C., Magorrian, J., \& Tremaine, S. 2000a,
  \aj, 119, 1157

\bibitem[Gebhardt et al.\ (2000b)]{gebhardt00b} Gebhardt, K., et
  al.\ 2000b, \apj, 539, L13

\bibitem[Gillessen et al.\ (2009)]{gill09} Gillessen, S., Eisenhauer,
  F., Trippe, S., Alexander, T., Genzel, R., Martins, F., \& Ott,
  T. 2009, \apj, 692, 1075 

\bibitem[Gomez \& Richtler (2004)]{gr04} Gomez, M. \& Richtler,
  T. 2004, A\&A, 415, 499

\bibitem[Goudfrooij et al.\ (2007)]{goudfrooij07} Goudfrooij, P.,
  Schweizer, F., Gilmore, D., \& Whitmore, B.C. 2007, \aj, 133, 2737 

\bibitem[Graham (2008)]{graham08} Graham, A.W. 2008, PASA, 25, 167

\bibitem[Graham et al.\ (2011)]{graham11} Graham, A.W., Onken, C.A.,
  Athanassoula, E., \& Combes, F. 2011, \mnras, 412, 2211

\bibitem[Graham \& Worley (2008)]{gw08} Graham, A.W. \& Worley,
  C.C. 2008, \mnras, 388, 1708

\bibitem[Gultekin et al.\ (2009)]{gultekin09} Gultekin, K., Richstone,
  D.O., Gebhardt, K., Lauer, T.R., Tremaine, S., Aller, M.C., Bender,
  R., Dressler, A., Faber, S.M., Filippenko, A.V., Green, R., Ho,
  L.C., Kormendy, J., Magorrian, J., Pinkney, J., \& Siopis, C. 2009,
  \apj, 698, 198

\bibitem[Hanes (1977)]{hanes77} Hanes, D.A. 1977, MmRAS, 84, 45
`
\bibitem[Hargis \& Rhode (2012)]{hargis12} Hargis, J.R. \& Rhode,
  K.L. 2012, \aj, in press

\bibitem[Hargis et al.\ (2011)]{hargis11} Hargis, J.R., Rhode, K.L.,
  Strader, J., \& Brodie, J.P. 2011, \apj, 738, 113 

\bibitem[Harris (1981)]{harris81} Harris, W. 1981, \apj, 251, 497

\bibitem[Harris (1991)]{harris91} Harris, W. 1991, \araa, 29, 543

\bibitem[Harris \& Harris (2011)]{hh11} Harris, G.L.H. \& Harris,
  W.E. 2011, \mnras, 410, 2347 

\bibitem[Harris et al.\ (2004)]{harris04} Harris, G.L.H., Harris,
  W.E., \& Geisler, D. 2004, \aj, 128, 723

\bibitem[Harris \& van den Bergh (1981)]{hvdb81} Harris, W.E. \& van
den Bergh, S. 1981, \aj, 86, 1627

\bibitem[Ho et al.\ (2009)]{ho09} Ho, L., Greene, J.E., Filippenko,
  A.V., \& Sargent, W.L.W. 2009, ApJSS, 183, 1

\bibitem[Hopkins et al.\ (2007)]{hopkins07} Hopkins, P.F., Hernquist,
  L., Cox, T.J., Robertson, B., \& Krause, E. 2007, \apj, 669, 67

\bibitem[Howard et al.\ (2008)]{howard08} Howard, C.D., Rich, M.R.,
  Reitzel, D.B., Koch, A., De~Propris, R., \& Zhao, H. 2008, \apj,
  688, 1060 

\bibitem[Hui et al.\ (1993)]{hui93} Hui, X., Ford, H.C., Ciardullo,
  R., \& Jacoby, G.H. 1993, \apj, 414, 463

\bibitem[Jahnke \& Maccio (2011)]{jm11} Jahnke, K. \& Maccio,
  A.V. 2011, \apj, 734, 92

\bibitem[Jardel et al.\ (2011)]{jardel11} Jardel, J.R., Gebhardt, K.,
  Shen, J., Fisher, D.B., Kormendy, J., Kinzler, J., Lauer, T.R.,
  Richstone, D., \& Gultekin, K. 2011, \apj, 739, 21 

\bibitem[Jarrett et al.\ (2000)]{jarrett00} Jarrett, T.H., Chester,
  T., Cutri, R., Schneider, S., Skrutskie, M., \& Huchra, J.P. 2000,
  \aj, 119, 2498

\bibitem[Jarrett et al.\ (2003)]{jarrett03} Jarrett, T.H., Chester,
  T., Cutri, R., Schneider, S.E., \& Huchra, J.P. 2003, \aj, 125, 525

\bibitem[Johansson et al.\ (2009)]{joh09} Johansson, P.H., Burkert,
  A., \& Naab, T. 2009, \apj, 707, L184

\bibitem[Jordan et al.\ (2007)]{jordan07} Jordan, A., McLaughlin,
  D.E., Cote, P., Ferrarese, L., Peng, E.W., Mei, S., Villegas, D.,
  Merritt, D., Tonry, J.L., \& West, M.J. 2007, ApJS, 171, 101

\bibitem[Kavelaars et al.\ (2000)]{kav00} Kavelaars, J.J., Harris,
  W.E., Hanes, D.A., Hesser, J.E., Pritchet, C.J. 2000, \apj, 533, 125 

\bibitem[Kent (1985)]{kent85} Kent, S.M. 1985, ApJS, 59, 115 

\bibitem[Kent (1988)]{kent88} Kent, S.M. 1988, \aj, 96, 2

\bibitem[Kormendy (1993)]{korm93} Kormendy, J. 1993, in Galactic
  bulges: proceedings of the 153rd Symposium of the International
  Astronomical Union, ed. H. DeJonghe \& H.J. Habing (Dordrecht:
  Kluwer), 209

\bibitem[Kormendy et al.\ (2011)]{korm11} Kormendy, J., Bender, R., \&
  Cornell, M.E. 2011, Nature, 469, 374

\bibitem[Kormendy \& Kennicutt (2004)]{kk04} Kormendy, J. \&
  Kennicutt, R.C., Jr. 2004, \araa, 42, 603

\bibitem[Kruijssen et al.\ (2012)]{kru12} Kruijssen, M.D., Inti
  Pelupessy, F., Lamers, H.J.G.L.M., Portegies Zwart, S.F., Bastian,
  N., Icke, V. 2012, \mnras, 421, 1927

\bibitem[Kundu \& Whitmore (2001)]{kw01} Kundu, A. \& Whitmore,
  B.C. 2001, \aj, 121, 2950

\bibitem[Kundu \& Zepf (2007)]{kz07} Kundu, A. \& Zepf, S.E. 2007,
  \apj, 660, L109 

\bibitem[Lee et al.\ (1998)]{lee98} Lee, M.G., Kim, E., \& Geisler,
  D. 1998, \aj, 115, 947

\bibitem[Mayer et al.\ (2010)]{mayer10} Mayer, L., Kazantzidis, S.,
  Escala, A., \& Callegari, S. 2010, Nature, 466, 1082

\bibitem[Miralda-Escude \& Kollmeier (2005)]{mk05} Miralda-Escude,
  J. \& Kollmeier, J. 2005, \apj, 619, 30

\bibitem[Moore et al.\ (2006)]{moore06} Moore, B., Diemand, J., Madau,
  P., Zemp, M., \& Stadel, J. 2006, MNRAS, 368, 563

\bibitem[Mortlock et al.\ (2011)]{mortlock11} Mortlock, S.J. et
  al.\ 2011, Nature, 474, 616

\bibitem[Muratov \& Gnedin (2010)]{muratov10} Muratov, A.L. \& Gnedin,
  O.Y. 2010, \apj, 718, 1266

\bibitem[Omukai et al.\ (2008)]{omukai08} Omukai, K., Schneider, R.,
  \& Haiman, Z. 2008, \apj, 686, 801

\bibitem[Peng et al.\ (2004)]{peng04} Peng, E.W., Ford, H.C., \&
  Freeman, K.C. 2004, ApJS, 150, 367

\bibitem[Peng et al.\ (2006)]{peng06} Peng, E.W., Jordan, A., Cote,
  P., Blakeslee, J.P., Ferrarese, L., Mei, S., West, M.J., Merritt,
  D., Milosavljevic, M., \& Tonry, J.L. 2006, \apj, 639, 95

\bibitem[Peng et al.\ (2008)]{peng08} Peng, E.W., Jordan, A., Cote,
  P., Takamiya, M., West, M.J., Blakeslee, J.P., Chen, C.-W.,
  Ferrarese, L., Mei, S., Tonry, J.L., \& West, A.A. 2008, \apj, 681,
  197

\bibitem[Perrett et al.\ (2002)]{perrett02} Perrett, K.M., Bridges,
  T.J., Hanes, D.A., Irwin, M.J., Brodie, J.P., Carter, D., Huchra,
  J.P., \& Watson, F.G. 2002, \aj, 123, 2490

\bibitem[Pinkney et al.\ (2003)]{pinkney03} Pinkney, J., Gebhardt, K.,
  Bender, R., Bower, G., Dressler, A., Faber, S.M., Filippenko, A.V.,
  Green, R., Ho, L.C., Kormendy, J., Lauer, T.R., Magorrian, J.,
  Richstone, D., \& Tremaine, S. 2003, \apj, 596, 903

\bibitem[Press et al.\ (1992)]{press92} Press, W. H., Teukolsky,
  S. A., Vetterling, W. T., \& Flannery, B. P. 1992, Numerical Recipes
  (2d ed.; Cambridge: Cambridge Univ. Press)

\bibitem[Rhode \& Zepf (2001)]{rz01} Rhode, K.L. \& Zepf,
S.E. 2001, \aj, 121, 210

\bibitem[Rhode \& Zepf (2003)]{rz03} Rhode, K.L. \& Zepf, S.E. 2003,
  \aj, 126, 2307

\bibitem[Rhode \& Zepf (2004)]{rz04} Rhode, K.L. \& Zepf,
S.E. 2004, \aj, 127, 302

\bibitem[Rhode et al.\ (2005)]{rzs05} Rhode, K.L., Zepf,
S.E., \& Santos, M.R.  2005, ApJ, 630, L21

\bibitem[Rhode et al.\ (2007)]{r07} Rhode, K.L., Zepf, S.E., Kundu,
  A., \& Larner, A.N. 2007, \aj, 134, 1403

\bibitem[Rhode et al.\ (2010)]{r10} Rhode, K.L., Windschitl, J.L., \&
  Young, M.D. 2010, \aj, 140, 430 

\bibitem[Robertson et al.\ (2006)]{rob06} Robertson, B., Hernquist,
  L., Cox, T.J., Di~Matteo, T., Hopkins, P.F., Martini, P., \&
  Springel, V. 2006, \apj, 641, 90

\bibitem[Sadoun \& Colin (2012)]{sadoun12} Sadoun, R. \& Colin,
  J. 2012, MNRAS, in press

\bibitem[Sandage (1961)]{sand61} Sandage, A. 1961, The Hubble Atlas
  of Galaxies (Washington, D.C.: Carnegie Inst.\ of Washington)

\bibitem[Santos (2003)]{santos03} Santos, M.R. 2003, in Extragalactic
Globular Cluster Systems, ed. M. Kissler-Patig (New York:
Springer-Verlag)

\bibitem[Schwarzschild et al.\ (1955)]{schwarz55} Schwarzschild, M.,
  Searle, L., \& Howard, R. 1955, \apj, 122, 353

\bibitem[Shen \& Gebhardt (2010)]{sg10} Shen, J. \& Gebhardt, K. 2010,
  \apj, 711, 484

\bibitem[Shen et al.\ (2010)]{shen10} Shen, J., Rich, R.M., Kormendy,
  J., Howard, C.D., De~Propris, R., \& Kunder, A. 2010, \apj, 720, L72 

\bibitem[Sheth (2003)]{sheth03} Sheth, R.K. 2003, MNRAS, 345, 1200

\bibitem[Simien \& de~Vaucouleurs (1986)]{sd86} Simien, F. \&
  de~Vaucouleurs, G. 1986, \apj, 302, 564

\bibitem[Snyder et al.\ (2011)]{snyder11} Snyder, G.F., Hopkins, P.F.,
  \& Hernquist, L. 2011, \apj, 728, L24

\bibitem[Spitler \& Forbes (2009)]{spitler09} Spitler, L.R. \& Forbes,
  D.A. 2009, \mnras, 392, L1

\bibitem[Spitler et al.\ (2008)]{spitler08} Spitler, L.R., Forbes,
  D.A., Strader, J., Brodie, J.P., \& Gallagher, J.S. 2008, \mnras,
  385, 361

\bibitem[Strader et al. (2007)]{strader07} Strader, J., Beasley, M.A.,
  \& Brodie, J.P. 2007, \aj, 133, 2015

\bibitem[Strader et al.\ (2004)]{strader04} Strader, J., Brodie, J.P.,
  \& Forbes, D.A. 2004, AJ, 127, 3431 

\bibitem[Sweigart \& Gross (1978)]{sweigart78} Sweigart, A.V. \&
  Gross, P.G. 1978, ApJS, 36, 405

\bibitem[Tamura et al.\ (2006)]{tamura06} Tamura, N., Sharples, R. M.,
  Arimoto, N., Onodera, M., Ohta, K., Yamada, Y. 2006, \mnras, 373,
  588 

\bibitem[Tremaine et al.\ (2002)]{tremaine02} Tremaine, S., Gebhardt,
  K., Bender, R., Bower, G., Dressler, A., Faber, S.M., Filippenko,
  A.V., Green, R., Grillmair, C., Ho, L.C., Kormendy, J., Lauer, T.R.,
  Magorrian, J., Pinkney, J., \& Richstone, D. 2002, \apj, 574, 740 

\bibitem[Tully (1988)]{tully88} Tully, R.B. 1988, Nearby Galaxies
  Catalog (New York: Cambridge Univ. Press)

\bibitem[van den Bergh (1993)]{vdb93} van den Bergh, S. 1993, \apj,
  411, 178

\bibitem[Vesperini (2000)]{vesper00} Vesperini, E. 2000, \mnras, 318, 841

\bibitem[Volonteri \& Rees (2006)]{volonteri06} Volonteri, M. \& Rees,
  M.J. 2006, \apj, 650, 669

\bibitem[Whitmore et al.\ (1995)]{whit95} Whitmore, B.C., Sparks,
  W.B., Lucas, R.A., Macchetto, F.D., \& Biretta, J.A. 1995, \apj,
  454, L73

\bibitem[Whitmore et al.\ (2010)]{whit10} Whitmore, B.C., Chandar, R.,
  Schweizer, F., Rothberg, B., Leitherer, C., Rieke, M., Rieke, G.,
  Blair, W.P., Mengel, S., \& Alonso-Herrero, A. 2010, \aj, 140, 75

\bibitem[Young et al.\ (2012)]{young12} Young, M.D., Dowell, J.L., \&
  Rhode, K.L. 2012, \aj, 144, 103

\bibitem[Zepf \& Ashman (1993)]{za93} Zepf, S.E., \& Ashman, K.M. 1993,
\mnras, 264, 611

\bibitem[Zinn (1985)]{zinn85} Zinn, R. 1985, \apj, 293, 424

\end{thebibliography}
\end{document}